\documentclass{IEEEoj}
\usepackage{cite}
\usepackage{amsmath,amssymb,amsfonts}
\usepackage{algorithmic}
\usepackage{graphicx,color}
\usepackage{textcomp}

\let\labelindent\relax
\usepackage{enumitem}
\usepackage{cite}
\usepackage{amsmath,amssymb,amsfonts}
\usepackage{algorithmic}
\usepackage{enumitem}
\usepackage{graphicx}
\usepackage{textcomp}
\usepackage{xcolor}
\usepackage{csquotes}
\usepackage{balance}
\usepackage{url}
\usepackage{balance}
\usepackage{rotating}
\usepackage{tikz} 
\usepackage{algorithm}

\usepackage{soul, color}
\usepackage[acronym,toc,shortcuts]{glossaries}
\usepackage{multirow}
\usepackage{comment}
\usepackage{soul}
\usepackage[hidelinks]{hyperref}

\begin{document}
\receiveddate{XX Month, XXXX}
\reviseddate{XX Month, XXXX}
\accepteddate{XX Month, XXXX}
\publisheddate{XX Month, XXXX}
\currentdate{11 January, 2024}

\title{Satellite Swarms for Direct-to-Cell Networks: A Distribution–Performance Trade-off Analysis}

\author{Xavier Artiga\IEEEauthorrefmark{1} , Marius Caus\IEEEauthorrefmark{1} \IEEEmembership{(Senior, IEEE)}, Ana I. Pérez-Neira\IEEEauthorrefmark{1,2}\IEEEmembership{(Fellow, IEEE)},

Yerassyl Akhmetkaziyev
 \IEEEauthorrefmark{3}, Malte Schellmann \IEEEauthorrefmark{3}, 
}
\affil{Centre Tecnològic de Telecommunicacions de Catalunya (CTTC/CERCA), Castelldefels (Barcelona), Spain}
\affil{Universitat Politècnica de Catalunya (UPC), Barcelona, Spain}
\affil{ Huawei Technologies German Research Center, Munich, Germany}
\corresp{CORRESPONDING AUTHOR: X. Artiga  (e-mail: xavier.artiga@cttc.es).}
\authornote{This work was supported by the project SOFIA PID2023-147305OB-C32 funded by MICIU/AEI/10.13039/501100011033 and by FEDER/UE}

\begin{abstract}
This work investigates 2D distributed satellite swarm configurations for direct-to-cell (D2C) applications. Unlike previous studies, which primarily considered swarms as deployment alternatives to monolithic arrays, this paper focuses on exploiting the increased spatial resolution enabled by large distributed apertures. \textcolor{black}{ Simulation results show that, for the considered scenarios, increasing the level of antenna distribution across satellite platforms enlarges the effective aperture and improves the system sum rate under ideal operating conditions.} While gains are moderate for uniformly distributed users, they become particularly pronounced in scenarios including hotspot with high user densities. The results further show that user scheduling strategies do not invalidate the superiority of highly distributed configurations, as these architectures provide a more balanced rate distribution across the coverage area, including hotspot regions. In addition, the paper analyzes several key implementation challenges associated with large distributed swarms, including errors in inter-satellite relative positioning, synchronization impairments, and limited beamforming and user-position update rates. \textcolor{black}{Although within the range of swarm configurations and scenarios considered in this work, performance gains continue to increase with aperture size}, these practical constraints may restrict swarm sizes in the medium term. Nevertheless, the observed performance gains  strongly motivate further research into scalable synchronization, positioning and data distribution techniques for future large-scale satellite swarms. 
\end{abstract}


\maketitle

\section{INTRODUCTION}
\label{secI}
The emergence of direct-to-cell (D2C) connectivity from Low-Earth Orbit  (LEO) satellites is widely recognized as one of the most transformative technological enablers envisioned for future 6G networks. By integrating non-terrestrial networks (NTNs) with conventional terrestrial infrastructures, D2C systems aim to provide truly ubiquitous connectivity, extending mobile coverage to virtually 100$\%$ of the globe, including remote, rural, maritime, and underserved regions. Despite the increasing momentum, currently deployed commercial D2C services remain limited in terms of capacity and supported use cases. Existing systems such as Starlink Direct-to-Cell, Beidou , Apple-Globalstar and Lynk \cite{bak25} primarily support low-data-rate applications, including emergency messaging and basic text communications. The main limiting factor is the extremely constrained link budget associated with handheld user equipment (UE), whose compact form factor imposes severe antenna gain limitations. Consequently, achieving broadband connectivity directly from satellites to conventional smartphones remains a major technical challenge.

Two main commercial strategies are being followed to mitigate the link budget limitations of D2C systems. The first approach, pursued by Starlink, focuses on reducing propagation losses through lower orbital altitudes while simultaneously increasing satellite transmit power by relaxing out-of-band emission constraints \cite{gar2025}. Nevertheless, these measures primarily provide incremental link budget improvements and come at the expense of additional challenges related to constellation density, onboard power consumption, interference mitigation, and satellite lifetime. A second and more aggressive strategy has been adopted by AST SpaceMobile, which relies on the use of large deployable spaceborne antennas to significantly increase the equivalent isotropically radiated power (EIRP) toward conventional handheld devices \cite{tuz23}, \cite{Ala24}. While this approach can substantially improve the downlink budget and potentially enable broadband connectivity for unmodified smartphones, it also raises critical concerns regarding scalability, deployment complexity, and overall satellite manufacturing costs.

As an alternative to monolithic large-aperture satellites, distributed satellite architectures have recently gained significant attention as a promising paradigm for future D2C  (see, e.g., \cite{bak25,sha26},  and the references therein). Instead of concentrating the communication capabilities into a single large and complex spacecraft, distributed approaches rely on smaller satellites operating in a coordinated manner. This paradigm offers several potential advantages, including improved scalability, increased redundancy, enhanced operational flexibility facilitating maintenance and replacement strategies, and reduced deployment risks. However, it needs to face fundamental challenges that  may  limit its practical performance or even feasibility. First, coherent cooperative transmission requires time, frequency and phase coherence among  the different satellite nodes, which not only implies tight synchronization at the transmitting satellites, but also at the receiving UEs to enable constructive and destructive addition of the signals coming from the different satellites. Although solutions for synchronization at transmitting satellites have been proposed \cite{Mar22}, differences in Doppler shifts and propagation times from the different satellites still may degrade the system performance. Second, the large distance between satellite nodes, well over the operating wavelength, results in grating lobes increasing potential interferences. The impact of these issues in the overall performance greatly depends on the  configuration of the distributed system, i.e. the number of nodes, distance between nodes and the antennas per node. In this context, two main approaches have emerged: using satellites from existing LEO constellations as distributed nodes, or deploying distributed antenna arrays formed by multiple satellites flying in a coordinated formation.

\subsection{Related work}
 The first distributed satellite approach considers satellites separated by large distances in order of tenths to hundreds of kilometres, typically selected from the same constellation, in a manner conceptually similar to coordinated multipoint (CoMP) or cell-free massive MIMO techniques in terrestrial communications  \cite{gaud25,xu24,and25,abd23,gui24,wan26,and24,xu23}. In this case, differences on the Doppler shift and propagation time from the different satellites dominate over synchronization errors, making the distributed approach infeasible unless a complex scheme of time and frequency pre-compensation is applied at each satellite and for each UE position \cite{gaud25,xu24,and25}. Note, however, that this scheme only aligns the signals at the intended user location. Consequently, interfering signals remain unaligned, limiting the applicability of interference-aware beamforming schemes such as Minimum Mean Square Error (MMSE) or Zero-Forcing (ZF). Despite this, many works neglect the misalignment of interfering signals. Under these ideal conditions, \cite{abd23} demonstrated the benefits of distributed satellite architectures over single-satellite and collocated massive MIMO solutions within a Time Division Duplexing (TDD) framework, where Channel State Information (CSI) is acquired through channel reciprocity from uplink UE transmissions. Assuming perfect interference alignment, \cite{gui24} resorted to Frequency Division Duplexing (FDD), which is generally more suitable for satellite communications due to the large propagation delays, combined with location-based CSI. The authors in \cite{wan26} also considered location-based CSI, referred to as statistical CSI, and assumed delay and Doppler pre-compensation while accounting for interference misalignment through a more accurate signal model. Specifically, the interfering signals transmitted from different satellites were assumed to be statistically independent, as the differences in propagation delays exceed the symbol period. It is worth noting that, even under this more realistic assumption, the satellites are unable to cooperate for interference mitigation. Alternative solutions to pre-compensation include resorting to multi-antenna user terminals capable of spatially separating the signals coming from different satellites \cite{and25}, though this is infeasible for compact handheld devices, or resorting to non-coherent satellite cooperation, though it provides lower performance \cite{and24}, \cite{sha26}. Regarding the grating lobes, due to the large distance between satellites, they fall within the main lobe of the individual satellite radiation patterns \cite{xu23}, \cite {xu24}. As a result, although power combination can still be achieved, the distributed array cannot effectively exploit an increased spatial resolution.  
 
 The second distributed satellite approach received much less attention so far. It consists of the use of fractionated satellites or satellite swarms, either tethered or flying in close formation, effectively forming a large distributed aperture. This concept was  originally proposed in the context of deep space exploration or communications \cite{had16}, but recently, it has been considered for D2C applications \cite{bac23,bac24,gaud25,tuz23,tuz23b,bei26,art25,tam26}.  In this configuration, the inter-satellite spacing is significantly reduced compared to widely distributed constellations, typically in the order of meters, which helps mitigate differential Doppler shifts. Differential propagation delays, although often neglected, remain as a relevant issue. To enable the use of narrowband beamforming models, either True Time Delay (TTD) compensation \cite{bac24} or Orthogonal Frequency Division Multiplexing (OFDM) signaling with subcarrier- (or subband-) dependent precoding must be adopted \cite{art25}. It is worth emphasizing that TTD compensation does not eliminate the interference misalignment problem. Grating lobes in this case do not affect the angular resolution but potentially increase interference levels \cite{gaud25,tuz23,art25}. In the framework of satellite swarms, two different schemes have been proposed. The first, nicknamed as Formation of Arrays (FoA), consists of few satellites nodes each of them equipped with large antenna arrays \cite{bac23},\cite{bac24,gaud25,bei26}.  In this case, grating lobes are partially mitigated by the radiation patterns of each satellite array. References \cite{bac23,gaud25} proposed for the LEO case to set the array size for each satellite according to the fairing size of the satellite launchers, whereas the number of satellites is constrained by a maximum aperture size providing a beam footprint around 3 km to resemble terrestrial cells. Remarkably, they concluded that such FoA scheme presents an antenna aperture similar to monolithic deployable antennas, questioning its usefulness. The alternative to FoA is the use of swarms of very small satellites equipped with a single antenna \cite{tuz23,tuz23b,art25,tam26} . In this case, grating lobes are a major issue and require specific swarm formation to reduce their levels. Authors in \cite{tuz23,tuz23b} proposed spiral based antennas, also adopted by \cite{tam26}, whereas we proposed random arrays \cite{art25}, which not only reduce grating lobes, but facilitate  mission operations by not requiring to maintain an specific formation with high accuracy. As in the case of \cite{gaud25}, in \cite{tuz23b} a fixed aperture diameter to synthesise a beam footprint around 3 km is also proposed. However, due to the single antenna arrangement of each satellite, its swarm counted with a very reduced number of total antennas, which translated to reduced total antenna gains. In both cases (i.e. \cite{gaud25} and \cite{tuz23b}), they evaluate system sum rate employing basic precoders like MMSE, Maximum Ratio Transmission (MRT) or ZF as a function of the number of uniformly distributed users. A clear optimum is always obtained, though its absolute value is different due to the EIRP differences of both approaches.  More recently, reference \cite{bei26} extended studies in \cite{tuz23,tuz23b} considering multiple antennas per satellite and a selective precoding architecture combining MRT in a first stage, and ZF or Signal-to-leakage-and-noise ratio precoder (SLNR) at beam level on a second stage, similarly to \cite{gui24}.

 A fair comparison between different configurations from monolithic arrays to single antenna satellite swarms going through FoA has not been carried out. It should be done in terms of cost–performance trade-off, which  requires expertise not only in wireless communications but also in aerospace engineering, orbital dynamics, and mission design. From previous studies in this line like \cite{Ala24} and \cite{had16}, it can be concluded that the suitability of fractionated architectures is highly mission-dependent. In  \cite{had16}, the use of distributed satellites is clearly advantageous in all aspects including cost reduction, whereas in \cite{Ala24} it provides benefits in terms of robustness, failure tolerance, and reduced impact on astronomical observations, but increases complexity of mission operations leading to higher operational costs. In any case, it is essential to assess the feasibility and performance from the communications point of view   before entering in detailed cost optimization analyses. Such analysis must include the sensitivity of the system performance to errors in synchronization, satellite position and attitude, and user position estimation, which deserve further attention. Authors in \cite{gaud25} analysed them for the GEO case with statistical model not directly connected to real position estimation errors. Similarly, \cite{bei26} also considered positioning errors from a statistical perspective, whereas \cite{bac24} just focused on the effects on the radiation patterns. Besides, \cite{tam26} studied robust precoding solutions against synchronization errors. Moreover, the performance analysis must go beyond the consideration of uniform user distributions, which have been the basis of all previous works, but do not reflect the reality \cite{fil24}. Only  \cite{gaud25} considered realistic non-uniform user distributions and scheduling, but limited it to the GEO case.
 
 \subsection{Novel Contributions} 
This work aims to fill these research gaps \textcolor{black}{by focusing on the satellite swarm concept.} The main goal is to answer the following question: \textit{\textcolor{black}{which level of antenna distribution provides the highest communication performance?}} To this end, we investigate \textcolor{black}{different satellite swarm configurations} under realistic user distributions, scheduling policies, and beamforming impairments. The main contributions of this paper \textcolor{black}{with respect to prior works addressing the satellite swarm concept}, illustrated in Table \ref{tab:comparison}, can be summarized as follows:

 \begin{table*}[!t]
 \color{black}
\centering
\caption{Novel contributions with respect to existing satellite swarm studies}
\label{tab:comparison}
\renewcommand{\arraystretch}{1.15}

\begin{tabular*}{\textwidth}{@{\extracolsep{\fill}}|l|c|c|c|c|c|c|c|c|}
\hline
\textbf{Topic} &
\textbf{\cite{tuz23,bac24}} &
\textbf{\cite{gaud25}} &
\textbf{\cite{bac23}} &
\textbf{\cite{tuz23b}} &
\textbf{\cite{bei26}} &
\textbf{\cite{art25}} &
\textbf{\cite{tam26}} &
\textbf{This work} \\
\hline

Evaluation of communication performance
 & $\times$ & \checkmark & \checkmark & \checkmark & \checkmark & \checkmark & \checkmark & \checkmark \\
\hline

Satellite distribution analysis
 & -- & $\times$ & $\times$ & $\times$ & $\times$ & $\times$ & $\times$ & \checkmark \\
\hline

Multicarrier/OFDM model with frequency-selective channel
& -- & $\times$ & $\times$ & $\times$ & $\triangle$ & $\times$ & $\triangle$ & \checkmark \\
\hline

Non-uniform user distribution
 & -- & $\triangle$ & $\triangle$ & $\times$ & $\times$ & $\times$ & $\times$ & \checkmark \\
\hline

Scheduling
 & -- & $\triangle$ & $\times$ & \checkmark & $\times$ & $\times$ & $\times$ & \checkmark \\
\hline

Sensitivity to positioning errors
 & -- & $\triangle$ & $\triangle$ & $\times$ & $\triangle$ & $\times$ & $\times$ & \checkmark \\
\hline

\end{tabular*}

\vspace{1mm}
\raggedright
\footnotesize
\textit{Note:} \checkmark~fully addressed; $\triangle$~partially addressed; $\times$~not addressed; --~not evaluated. For works that  do not evaluate communication performance, the remaining entries are marked as not evaluated.
\end{table*}

\begin{itemize}
\color{black}
\item A fair comparison of antenna distribution architectures, ranging from monolithic large-aperture satellites to FoAs with different numbers of antennas per satellite and fully distributed single-antenna satellite swarms, under a fixed EIRP. To the best of the authors' knowledge, this analysis has not been previously reported. Existing works perform, at most, a comparison with a single monolithic array (e.g., \cite{tam26}).

\item In-depth analysis of multicarrier and OFDM signal models for distributed satellite swarms, deriving the conditions under which wideband beamforming in the presence of differential propagation delays can be decomposed into independent narrowband problems across subcarriers. Consequently, all simulations are performed using a multicarrier model. Existing works either consider TTD compensation without analyzing interference misalignment \cite{bac23,bac24,gaud25}, adopt a narrowband model without justification \cite{tuz23b}, or employ OFDM without discussing the impact of differential propagation delays or providing evidence that the resulting frequency-selective channel is modelled, rather than implicitly assuming frequency-flat propagation \cite{art25,bei26,tam26}.

\item Analysis of the interplay between array configuration and user spatial distribution, covering uniform deployments, single-hotspot, and mixed multi-cluster scenarios. Previous studies primarily assume uniformly distributed users \cite{art25,bei26,tuz23b,bac23}; \cite{bac23} additionally considers isolated hotspots matching the beam footprint but allocates users to orthogonal time/frequency resources; \cite{gaud25} enforces a minimum inter-user distance through an assumed scheduling and \cite{tam26} does not specify the user distribution but only considers 10 users. 

\item Evaluation of the impact of inter-satellite spacing and its relationship with hotspot size. Although \cite{bac24} studies inter-satellite spacing through array thinning strategies, the analysis is restricted to radiation pattern characteristics, without evaluating the resulting system capacity or its interplay with the user spatial distribution. 

\item A minimum-distance user scheduling algorithm that, unlike conventional partition-based schedulers proposed in \cite{tuz23b} and \cite{gaud25} (the latter only for the GEO case), generates overlapping co-scheduled user groups, allowing users to be served in multiple scheduling resources while enforcing a minimum separation between users sharing the same resource. This significantly improves the overall system throughput, particularly for sparsely distributed users.


\item Sensitivity analysis to phase errors caused by satellite position uncertainties, satellite-user relative mobility, and user location update periodicity using a physics-based geometric model of the relative satellite-user motion. Unlike \cite{bei26}, which considers only user positioning errors modelled as a Gaussian random variable, and \cite{bac23,gaud25}, which represent multiple error sources through a uniformly distributed phase term, the proposed model explicitly captures the underlying geometry and mobility driving the phase errors

\item Identification of tracking errors as the primary factor limiting the maximum effective aperture size from a performance perspective. In contrast to \cite{bac23,gaud25,tuz23b}, where swarm size is constrained by typical terrestrial cell dimensions without a dedicated system-level analysis, this work identifies the performance-driven aperture limit imposed by tracking errors.
\end{itemize}

\subsection{Paper Organization}
The remainder of the paper is organized as follows. Section \ref{secII} presents an error-free system model; \textcolor{black}{Section \ref{secIII}  presents the theoretical framework for predicting the radiation patterns of satellite swarms and relates them to the array characteristics of representative swarm configurations}; Section \ref{secIV} presents the beamforming schemes used in the evaluation; Section \ref{secV} introduces the considered user distributions and the proposed scheduling solution; Section \ref{secVI} describes the phase error model considering imperfect estimation of the relative position between satellite nodes, as well as imperfect estimation of the user position due to limited user position and beamformer update rates; and Section \ref{secVII} presents the numerical results. Finally, Section \ref{secVIII} concludes the work.

\section{System model}
\label{secII}

\begin{figure}
    \centering
    \includegraphics[width=0.45\textwidth]{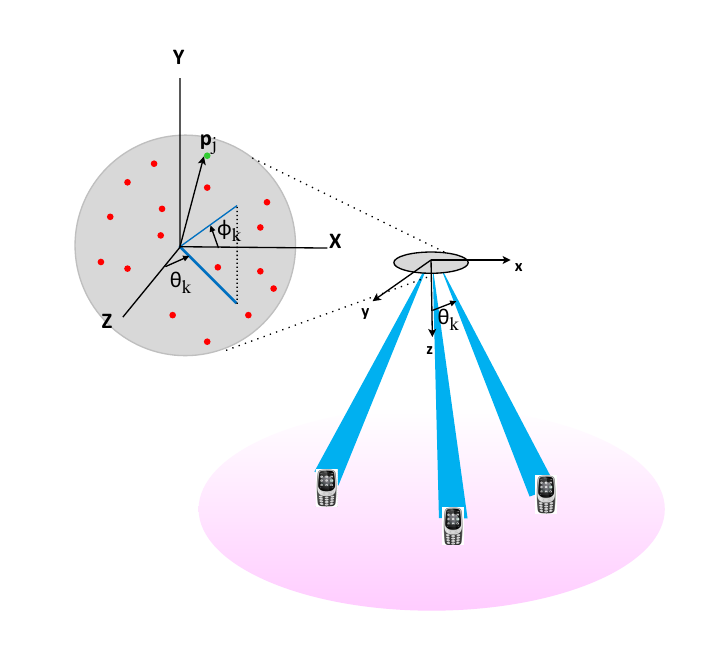}
    \caption{Illustration of a multi-beam satellite swarm communication system.}
    \label{fig1}
\end{figure}

This section presents the system model consisting of either a monolithic satellite or a distributed satellite swarm synthesising multiple beams towards \textit{K} target users in the ground. The swarm case considers $N_s$ satellite nodes in a formation flying configuration, i.e. with inter-node distances in the order of meters, performing a cooperative transmission.  The concept is illustrated in Figure \ref{fig1}. Independently of the specific satellite configuration, it is assumed that a planar array  with \textit{N}  radiating elements is formed in the space segment. Such array is controlled by a central node that implements the feeder link towards the gateway ground station and calculates beamforming weights in a centralised manner. Data distribution from a central node to all radiating elements is a key challenge in any massive MIMO system. It has been studied in terrestrial systems (see, for instance, \cite{11177239} and references therein), where partially and fully decentralized solutions have been proposed. The adoption of these solutions in satellite systems remains an open issue. However, these aspects are outside the scope of this paper, which focuses on space-to-Earth transmission performance, with particular emphasis on the downlink. Therefore, ideal data distribution is assumed. In addition, ideal synchronization among all radiating elements is initially assumed, although Section \ref{secVI} accounts for phase errors due to imperfect synchronization and imperfect estimation of the positions of both the distributed satellite nodes and the target users.      

For convenience, the coordinate reference system is defined so that the array lies in the XY plane. the coordinates of the \textit{n}-th antenna are arranged in the vector
\begin{equation}
\mathbf{p}_n=\left[d_n \cos \varphi_n , d_n \sin \varphi_n, 0\right],    
\label{eq1}\end{equation}
for $1\leq n\leq N$. The elements of the array are confined within a circle of radius $R$; hence $d_n \leq R$. At a given time instant, we assume that the array synthesizes $K$ beams in the direction of the users. The triplet $(r_k,\phi_k,\theta_k)$ defines the spherical coordinates of the \textit{k}-th terminal. 

\subsection{Transmitter}
\label{secIIa}
The signal that is transmitted by the \textit{n}-th radiating element can be formulated as
\begin{equation}
s_n (t)= \sum_i p(t-iT_s)\sum_{k=1}^{K} b_n \left(\theta_k,\phi_k\right) x_k[i].   
\label{eq2}\end{equation}
The term $b_n \left(\theta_k,\phi_k\right)$ corresponds to the \textit{n}-th antenna coefficient that weights $x_k[i]$, which is the information symbol intended to the \textit{k}-th user. It shall be noted that \textit{K} users are scheduled in the same time and frequency resources. 

At transmission, the modulated signal is obtained by shaping the precoded symbols with the pulse $p(t)$, which is defined for the symbol period $T_s$. The use of square root raised cosine pulses is common, due to its remarkable properties, such as good spectral confinement and perfect reconstruction.

\subsection{Satellite channel model}
\label{secIIb}
The channel impulse response between the \textit{k}-th user and the \textit{n}-th antenna is represented by $h_n \left(t,\tau,\theta_k,\phi_k \right)$. To simplify the model, the channel is defined for an interval where $\left(\theta_k,\phi_k \right)$ remains constant. Under these premises, the channel response can be written as
\begin{equation}\begin{array}{rl}
h_n \left(t,\tau,\theta_k,\phi_k \right)=&a_n \left(\theta_k,\phi_k \right) \delta \left(\tau-\tau_n\left(\theta_k,\phi_k \right)\right)\times\\
& e^{j2\pi f_D \left(\theta_k,\phi_k \right) \left(t-\tau_n\left(\theta_k,\phi_k \right) \right)},
\end{array}\label{eq4}\end{equation}
where $a_n \left( \theta_k,\phi_k\right)$ is the channel gain, $\tau_n\left( \theta_k,\phi_k\right)$ is the differential delay relative to the \textit{n}-th radiating element and $f_D \left(\theta_k,\phi_k\right)$ is the Doppler shift associated with the direction $\left(\theta_k,\phi_k \right)$. Under the assumption of a monolithic satellite or of satellite nodes in close proximity, then it follows that Doppler shifts for different paths are almost identical. Considering the methodology described in \cite{nr811}, the Doppler shift is assumed to remain constant across the signal bandwidth. The approximation holds when the carrier frequency is significantly higher than the signal bandwidth, i.e., \textcolor{black}{$B_W\ll f_c$}. 

Taking as reference the origin of the coordinate reference system, the differential delay is defined by 
\begin{equation}
\begin{array}{rl}
\tau_n \left(\theta_k,\phi_k \right)=& \frac{1}{c} \left( d_n \sin \theta_{k} \cos \left(\phi_k-\varphi_n \right)\right) \\
=&\frac{1}{c} \left( d_n \cos \varphi_n u_k^x+d_n \sin \varphi_n v_k^y\right),
\end{array}\label{eq5}\end{equation}
where $c$ denotes the speed of light and $u_k^x=\sin \theta_{k} \cos \phi_k$, $v_k^y=\sin \theta_{k} \sin \phi_k$ are the angular coordinates of the \textit{k}-th user.
\color{black}
The channel coefficient $a_n(\theta_k,\phi_k)$ is modelled as a Rician random variable, composed of a deterministic line-of-sight (LoS) component, accounting for the direct propagation path, and a statistically modelled diffuse non line-of-sight (NLoS) component representing multipath propagation. It reads as

\begin{equation}
\begin{aligned}
a_n(\theta_k,\phi_k)
=&\,
A_n(\theta_k,\phi_k)
\Bigg(
\sqrt{\frac{K_R}{K_R+1}}
e^{j\phi_n(\theta_k,\phi_k)}
\\
&\qquad\qquad+
\sqrt{\frac{1}{K_R+1}}
\xi_k
\Bigg),
\end{aligned}
\label{eq:rice}
\end{equation}
where $K_R$ denotes the Rician factor,
$A_n(\theta_k,\phi_k)$ is the large-scale channel amplitude, and
$\phi_n(\theta_k,\phi_k)$ is the deterministic propagation phase. The large-scale channel amplitude is given by


\begin{equation}
A_n(\theta_k,\phi_k)
=
\sqrt{
\frac{G_n(\theta_k,\phi_k)G_R}
{PL(\theta_k)OL}
},
\label{eq:An}
\end{equation}
whereas the deterministic propagation phase is

\begin{equation}
\phi_n(\theta_k,\phi_k)
=
\frac{2\pi f_c}{c}r(\theta_k)
-
2\pi f_c\tau_n(\theta_k,\phi_k).
\label{eq:phase}
\end{equation} The large-scale channel amplitude depends on the terminal antenna gain $G_R$, assumed isotropic, the gain of the \textit{n}-th radiating element $G_n \left(\theta_k,\phi_k\right)$, the path loss $\text{PL}(\theta_k)$ and other losses $\text{OL}$. In clear sky conditions, the path loss can be defined in dB as
\begin{equation}
\text{PL}(\theta_k)=32.45+20\log_{10} f_c+20\log_{10} r(\theta_k),
\label{eq7}\end{equation}
where $f_c$ is expressed in GHz and $r(\theta_k)$ denotes the slant range in meters along the direction of the \textit{k}-the user. The relation between the slant range and the angle of departure $\theta_k$ is described in \cite{ang20}. Other losses OL are incorporated to account for miscellaneous attenuations, including implementation losses, power amplifier back-off and/or additional propagation impairments.

The diffuse component is modelled by a complex fading coefficient
$\xi_k$, assumed to be a zero-mean circularly symmetric complex Gaussian
random variable $\xi_k \sim \mathcal{CN}(0,1)$. Remarkably, all the antenna elements experience the same path loss and other losses, due to the fact that the inter-satellite distance is low with respect to the slant range. This also applies to the diffuse component, which is totally correlated among distributed nodes.

Under this model, the NLoS component does not modify the spatial signature of
the user channel and therefore does not alter the beamforming patterns nor the
trade-offs between beamwidth, sidelobe levels, and aperture size analysed in
this work. Instead, it introduces random fluctuations of the received signal
power while preserving the average channel power. Consequently,  the instantaneous SINR exhibits a
larger variance than in the pure LoS case. Since the achievable rate is a
concave function of the SINR, this increased variability leads to a 
reduction of the ergodic sum rate. For the relatively high Rician factors considered (e.g., $K_R=$10), this reduction is moderate and does not modify the qualitative conclusions regarding the relative performance of the different swarm configurations. 
\color{black}

\subsection{Receiver}
\label{secIIc}
According to the model formulated in Section \ref{secIIb}, the signal received by user $k$ is given by
\begin{equation}\begin{array}{rl}
r_k(t)=&\displaystyle \sum_{n=1}^{N} \int_\tau h_n \left(t,\tau,\theta_k,\phi_k \right) s_n(t-\tau)  \partial \tau +z_k (t)\\
=&\displaystyle \sum_{n=1}^{N} a_n \left(\theta_k,\phi_k \right) s_n \left(t-\tau_n \left(\theta_k,\phi_k \right)\right)\times \\
& \displaystyle e^{j2\pi f_D \left(\theta_k,\phi_k \right) \left(t-\tau_n\left(\theta_k,\phi_k \right) \right)}+z_k (t),
\end{array}\label{eq8}\end{equation}
for $1\leq k \leq K$. Let $z_k (t)$ denote the noise that contaminates the reception of the \textit{k}-th terminal. It follows a white Gaussian stochastic model with zero mean and variance $\sigma^2=K_B T B_W $, denoting $K_B$ the Boltzmann constant , $B_W$ the bandwidth and $T$ the equivalent noise temperature of the receiver\footnote{$T=T_A+(F-1)T_0$, with $T_A$ denoting antenna temperature, F the noise factor and $T_0$ the reference temperature}. 

In alignment with 3GPP requirements, it is assumed that the UE is aware of both the satellite/swarm trajectory and its own location, as specified in \cite{nr300}. With this information, the terminal shall be able to autonomously compensate for the Doppler frequency shift. After applying the compensation mechanism, it can be readily verified that the received signal becomes

\begin{equation}\begin{array}{rl}
r^c_k(t)=\displaystyle \sum_{n=1}^{N} a_n \left(\theta_k,\phi_k \right) s_n \left(t-\tau_n \left(\theta_k,\phi_k \right)\right)+z^c_k (t).
\end{array}\label{eq9}\end{equation}
It is important to remark that the stochastic model of the noise is not altered after performing the frequency correction. In the next stage, the matched filter is applied to recover the transmitted symbols, leading to 
\begin{equation}
y_k(t)=r^c_k(t)\star p^*(-t)=\int_\tau r^c_k(\tau)p^*(\tau-t) \partial \tau.    
\label{eq10}\end{equation}
Sampling the output every $T_s$ seconds, the received sequence becomes
\begin{equation}
 \color{black}
\begin{array}{rl}
\hat{x}_k[i]=&\displaystyle \sum^{K}_{j=1}\sum^{N}_{n=1}a_n \left(\theta_k,\phi_k \right) b_n \left(\theta_j,\phi_j\right)\times \\
&\displaystyle\sum_l R_p\left(lT_s-\tau_n \left(\theta_k,\phi_k \right)  \right)x_j[i-l]\\
&+w_k[i],
\end{array}\label{eq11}\end{equation}
where $R_p(t) = p(t) \star p^*(-t)$. Assuming a Nyquist orthogonal transmission, then it follows that the noise samples are independent and identically distributed (i.i.d.) random Gaussian variables, i.e., $w_k[i] \sim \mathcal{CN}\left(0, \sigma^2\right)$. Due to the fact that the differential delay between satellites is not negligible, the symbols are impaired by inter-symbol interference (ISI) in addition to inter-user interference (IUI). 

\subsection{Narrowband conditions}
Under narrowband conditions, the magnitude of the delays is such that \textcolor{black}{$R_p\left(lT_s-\tau_n \left(\theta_k,\phi_k \right)\right) \approx R_p\left(lT_s \right)$, $\forall n$}. In this case, we can invoke the Nyquist orthogonal transmission property to express the end-to-end communication system as 
\begin{equation}\begin{array}{rl}
\hat{x}_k[i]=&\displaystyle \sum^{K}_{j=1}\sum^{N}_{n=1}a_n \left(\theta_k,\phi_k \right) b_n \left(\theta_j,\phi_j\right)x_j[i]+w_k[i].
\end{array}\label{eq12}\end{equation}
Adopting a matrix notation, the input-output relation can be compactly expressed as 
\begin{equation}\begin{array}{rl}
\hat{x}_k[i]=&\displaystyle \sum^{K}_{j=1}\mathbf{a}^\mathrm{T}_k \mathbf{b}_j x_j[i] + w_k[i],
\end{array}\label{eq14}\end{equation}
where
\begin{equation}
 \mathbf{a}_k=\left[a_1 \left(\theta_k,\phi_k\right), \cdots, a_{N} \left(\theta_k,\phi_k\right) \right]^\mathrm{T} \in \mathbb{C}^{N\times 1}   
\label{eq13}\end{equation}
\begin{equation}
 \mathbf{b}_k=\left[b_1 \left(\theta_k,\phi_k\right), \cdots, b_{N} \left(\theta_k,\phi_k\right) \right]^\mathrm{T}  \in \mathbb{C}^{N\times 1}. 
\label{eq13b}\end{equation}
It is worth noting that the system that results from the narrowband conditions, offers better analytical tractability than the general model. The key aspect stems from the fact that ISI can be neglected. This condition is satisfied if 
\begin{equation}
\operatorname{max}_n \tau_n\left( \theta_k, \phi_k \right) \leq \beta T_s,    
\label{eq15}\end{equation}
where \textcolor{black}{$\beta\ll1$}. Consequently, the model expressed in (\ref{eq14}) can be adopted without any significant modelling error when the maximum differential delay is significantly lower than the symbol period.

In systems where (\ref{eq15}) is not satisfied, ISI can still be mitigated. A viable approach consists in pre-compensating the differential delay for each user at the transmitter \cite{bac24}. This scheme, which is referred to as True Time Delay (TTD), aims to ensure that signals are received synchronously in the target direction. This can be achieved by implementing an adjustable time offset on each antenna element. Assuming perfect delay compensation, the received signal can be recast as

\begin{equation}
\color{black}
\begin{array}{c}
\hat{x}_k[i]=\mathbf{a}^\mathrm{T}_k \mathbf{b}_k x_k[i] +\displaystyle \sum_{j\neq k}\sum^{N}_{n=1}a_n \left(\theta_k,\phi_k \right) b_n \left(\theta_j,\phi_j\right) \times \\
\displaystyle \sum_l R_p\left(lT_s-\tau_n \left(\theta_k,\phi_k \right) +\tau_n \left(\theta_j,\phi_j \right) \right) x_j[i-l]+ w_k[i].
\end{array}\label{eq16}\end{equation}
This expression reveals that the desired signals are aligned, while the interfering signals exhibit different arrival times. This observation suggests that the TTD can be readily combined with beamforming designs that aim to maximise the received signal energy. However, incorporating interference mitigation into the beamforming design becomes challenging. The main difficulty lies in handling the large number of interfering components arising from the lack of synchronization between desired and unwanted signals.   

Another approach to eliminate ISI is based on implementing a multicarrier modulation scheme, so that the symbol period is scaled by the number of subcarriers $N_{sc}$. Hence, if $N_{sc}$ is sufficiently large, the narrowband condition will be satisfied at the subcarrier level. The resulting symbol period can be formulated as $T_{s,mc}=(1+\alpha)\frac{N_{sc}}{B_W}$, where $\alpha$ is the roll-off factor. It shall be noted that the system adopts a multicarrier architecture, in which individual carriers are frequency-separated with guard bands to avoid inter-carrier interference (ICI). This configuration is compatible with DVB-S2 systems, in which the available bandwidth can be partitioned into multiple independent subcarrier signals, each occupying a portion of the satellite transponder bandwidth. 

To determine the number of subcarriers required to satisfy the narrowband condition, we focus on deriving the maximum differential delay. For analytical tractability, we assume that the satellite nodes are enclosed within a circle of diameter $2R$. Under this topology, the maximum differential slant range within the swarm can be expressed as
\begin{equation}
 \Delta SR (\theta) = 2R \sin \theta.  
\label{eq17}\end{equation}
From this expression, it is straightforward to compute the corresponding differential delay as $\Delta SR (\theta)/ c$. For a given symbol period $T_{s,mc}$, the dimensionality constraint that stems from (\ref{eq15}) becomes 
\begin{equation}
2R \leq  \frac{\beta T_{s,mc} c}{\sin \theta}.
\label{eq18}\end{equation}
Remarkably, the maximum steering angle $\theta_{\text{max}}$ determines the maximum swarm diameter. Typically, $\theta_{\text{max}}$ is derived from the minimum elevation angle at which terminals can establish communication with the swarm. In multicarrier systems where subcarrier signals do not overlap in frequency, the minimum number of subcarriers required to satisfy the narrowband condition can be expressed as 
\begin{equation}
\frac{B_W}{(1+\alpha)}\frac{2R   \sin \theta_{\text{max}}}{\beta c} \leq  N_{sc}.
\label{eq19}\end{equation}
When this condition holds, the narrowband model in (\ref{eq14}) can be used to characterise the input-output relation on each subcarrier. To illustrate with a numerical example, we consider a satellite swarm characterised by $R=30$ m, $B_W=20$ MHz, $\theta_{\text{max}}=44.4^\circ$ and $\alpha=0.1$. For $\beta=0.05$, the minimum number of subcarriers required to satisfy the narrowband assumption is by $N_{sc}=51$. 

To ensure compliance with 3GPP specifications, it is necessary to adopt a multicarrier method compatible with OFDM. Under this framework, the narrowband assumption holds if the cyclic prefix (CP) duration exceeds the maximum channel delay. Since in the OFDM scheme the parameter $N_{sc}$ corresponds to the fast Fourier transform (FFT) size and the useful symbol duration is $N_{sc}/B_W$, the narrowband condition can be written as
\begin{equation}
\frac{B_W 2R \sin  \theta_{\text{max}}}{\rho c} \leq N_{sc},   
\end{equation}
where $\rho$ denotes the CP ratio. A typical value for the normal CP is $\rho=\frac{1}{14}$. Given this configuration, and using the same parameters as in the previous numerical example, the minimum number of OFDM subcarriers is given by $N_{sc}=40$. Clearly, this shows that in OFDM systems, the condition required to justify the narrowband model is less stringent. 

\subsection{OFDM}
In OFDM systems, all the processing is performed on a per-subcarrier basis. Consequently, the signal demodulated by the \textit{k}-th user on the \textit{q}-th subcarrier is defined by 
\begin{equation}
\hat{x}_{k,q}[i]=\sum^{K}_{j=1}\mathbf{a}^\mathrm{T}_{k,q} \mathbf{b}_{j,q} x_{j,q}[i] + w_{k,q}[i].   
\end{equation}
For an even number of subcarriers $N_{sc}$, the subcarrier index is defined over the interval $-\frac{N_{sc}}{2}\leq q\leq \frac{N_{sc}}{2}-1$. Considering the channel model introduced in subsection \ref{secII}.\ref{secIIb}, let $\mathbf{a}_{k,q}$ denote the channel vector in (\ref{eq13}) evaluated at the subcarrier frequency $f_q=f_c+q\frac{B_W}{N_{sc}}$. 

\subsection{Performance Metrics}
Assuming the use of OFDM, the SINR at subcarrier \textit{q} is given by
\begin{equation}
\color{black}
\mathrm{SINR}_{k,q}=
\frac{|\mathbf{a}_{k,q}^\mathrm{T} \mathbf{b}_{k,q}|^2}{\displaystyle\sum_{j \neq k}|\mathbf{a}_{k,q}^\mathrm{T} \mathbf{b}_{j,q}|^2 + \frac{\sigma^2}{N_{sc}}}.
\label{eq20c}\end{equation}
In (\ref{eq20c}), an equal power allocation per subcarrier is assumed, so the power of the beamformer $\mathbf{b}_{k,q}$ also scales down with the number of subcarriers, i.e. $|\mathbf{b}_{k,q}|^2=\frac{|\mathbf{b}_{k}|^2}{N_{\mathrm{sc}}}$ denoting $\mathbf{b}_{k}$ de beamformer applied to the whole band. In this framework, the user rate in the Shannon sense can be defined by
\begin{equation}
\color{black}
R_{k,q}=
\begin{cases}
\frac{f_{\mathrm{sch},k}}{1+\rho}\frac{B_W}{N_{sc}}\,\log_2(1+\mathrm{SINR}_{k,q}),
& \mathrm{SINR}_{k,q} \geq \mathrm{SINR}_{\mathrm{th}}
\\[6pt]
0,
& \mathrm{SINR}_{k,q} < \mathrm{SINR}_{\mathrm{th}}.
\end{cases}
\label{eq21}\end{equation}
\textcolor{black}{where $f_{\mathrm{sch},k}\in[0,1]$ denotes the scheduling factor of user $k$, defined as the fraction of transmission slots during which the user is scheduled.} 
Finally the system sum rate writes as 
\begin{equation}
\mathrm{SR}=\sum^{K}_{k=1}\sum^{\frac{N_{sc}}{2}-1}_{q=-\frac{N_{sc}}{2}}R_{k,q}.
\label{eq22}\end{equation}


\section{Swarm model}
\label{secIII}
\subsection{Theoretical background}
\color{black}
Equations \eqref{eq21} and \eqref{eq22} indicate that the system sum-rate depends on the users' SINR, which, in multibeam satellite systems, is mainly limited by inter-beam interference, particularly by the interference originating from the main lobe of adjacent beams. This has motivated the adoption of different transmission strategies, such as frequency reuse schemes, interference-aware precoding/beamforming, and minimum-distance user scheduling \cite{ang20}. This work complements existing interference mitigation techniques by increasing the spatial resolution of the antenna array through the distribution of the transmitting antennas over multiple satellites.

For sufficiently uniform satellite distributions, the main beam characteristics are primarily determined by the effective aperture rather than by the exact element locations. For a uniformly illuminated circular aperture of diameter $D$, the half-power beamwidth can be accurately approximated by \cite{balanis}
\begin{equation}
\theta_{\mathrm{HPBW}}
\approx
50.8
\frac{\lambda}{D}
\qquad [^\circ].
\label{eq:HPBWdeg}
\end{equation}
showing that the angular resolution is mainly governed by the aperture diameter.

Distributing the radiating elements over multiple satellites results in average inter-element spacings much larger than the operating wavelength. For periodic layouts, this would produce severe grating lobes. A well-established approach to mitigate this effect is to randomize the satellite positions. Still, a potential trade-off between the beamwidth reduction and the distribution and level of sidelobes may arise. To address it, we must refer to the array radiation pattern, which can be expressed as the product of three contributions,
\begin{equation}
P(\theta,\phi)=
E(\theta,\phi)\,
AF_{\mathrm{sub}}(\theta,\phi)\,
AF_{\mathrm{rand}}(\theta,\phi),
\label{eq:pattern}
\end{equation}
where $E(\theta,\phi)$ denotes the individual radiating element pattern, $AF_{\mathrm{sub}}$ represents  array factor of each satellite subarray, and $AF_{\mathrm{rand}}$ accounts for the array factor of the distributed satellite centroids.

Without loss of generality, it is considered that each satellite equips a uniform planar square subarray with $N_{\mathrm{sub}}$ elements, therefore the subarray factor is  given by \cite{balanis}
\begin{equation}
AF_{\mathrm{sub}}
=
\frac{\sin\left(\sqrt{N_{\mathrm{sub}}}\psi/2\right)}
{\sqrt{N_{\mathrm{sub}}}\sin\left(\psi/2\right)}
\,
\frac{\sin\left(\sqrt{N_{\mathrm{sub}}}\psi/2\right)}
{\sqrt{N_{\mathrm{sub}}}\sin\left(\psi/2\right)},
\end{equation}
with
\[
\psi_x=k_0 d_s u,\qquad
\psi_y=k_0 d_s v,
\]
where $u=sin(\theta)cos(\phi)$; $v=sin(\theta)sin(\phi)$;  $k_0$ is the free-space wavenumber and $d_s$ is the inter-element distance in the satellite subarray. This factor  exhibits larger sidelobe levels (SLL) in the two principle planes of the subarray with  peak SLL of approximately $-13.26$ dB.

 The random array factor can be generally expressed as

\begin{equation}
AF_{\mathrm{rand}}(\mathbf{q})
=
\frac{1}{N_s}
\sum_{n_s=1}^{N_s}
e^{j\mathbf{q}\cdot\mathbf{p}_{ns}},
\end{equation}
where $\mathbf{q}=k_0[u,v]^T$  is the spatial-frequency (wavevector) associated with the observation direction, $N_s$ denotes the number of distributed satellites and $\mathbf{p}_{ns}$ the satellite positions. The latter corresponds to \eqref{eq1} but referring only to satellite centroid positions instead of radiating element positions.



The statistical behaviour of the random array factor can be analysed using the recently proposed extension of the classical random-array theory in \cite{christian2026}. Unlike the original formulation in \cite{lo1964}, which assumes independently distributed radiating elements, this model explicitly accounts for the spatial correlations introduced by the minimum-distance constraint through the so-called structure factor $S(\mathbf{q})$ associated with the underlying point process of antenna distribution. Note that, in distributed satellite swarms, this minimum distance naturally corresponds to the minimum separation that neighbouring satellites can safely maintain.

The expected normalized random-array power pattern is therefore given by \cite{christian2026}

\begin{equation}
\mathbb{E}\!\left\{|AF_{\mathrm{rand}}(\mathbf{q})|^2\right\}
=
\left(1-\frac{1}{N_s}\right)
|\Phi(\mathbf{q})|^2
+
\frac{S(\mathbf{q})}{N_s}
-
\frac{S(\mathbf{q})-1}{N_s^2}.
\label{eq:christian}
\end{equation}
The first term represents the coherent contribution associated with the average aperture illumination, while the remaining terms correspond to the incoherent contribution arising from the random satellite locations. Since the satellite centroids are approximately uniformly distributed over a circular aperture, the average array factor is the Fourier transform of a uniformly illuminated circular disk,
\begin{equation}
\Phi(\mathbf{q})
=
\mathbb{E}\!\left\{AF_{\mathrm{rand}}(\mathbf{q})\right\}
=
\frac{2J_1(|\mathbf{q}|R)}
{|\mathbf{q}|R},
\qquad
R=\frac{D}{2},
\end{equation}


Unlike the classical random-array formulation, where the incoherent contribution is spatially uniform \cite{lo1964}, the structure factor modulates the sidelobe floor as a function of the spatial frequency. Consequently, the sidelobe statistics become angle dependent and directly reflect the spatial correlations introduced by the minimum-distance constraint.

For a hard-core point process with exclusion radius $r_{\min}$, \cite{christian2026} derived the following closed-form approximation

\begin{equation}
S(|\mathbf{q}|)
=
1-
2\pi\nu r_{\min}
\frac{J_1\!\left(|\mathbf{q}| r_{\min}\right)}
{|\mathbf{q}|},
\label{eq:Sk}
\end{equation}
where $\nu = \frac{4N_s}{\pi D^2}$ is the spatial density of satellite centroids. This closed-form expression predicts the existence of a low-energy region around the origin of the spatial spectrum. In particular, the term
$\frac{J_1\!\left(|\mathbf{q}| r_{\min}\right)}
{|\mathbf{q}|}.$ reaches its maximum value at low spatial frequencies and decreases monotonically until the first zero of $J_1(\cdot)$. Since this positive contribution is subtracted from unity in (\ref{eq:Sk}), the structure factor remains below a Poisson reference level over a finite region around the origin, indicating that the incoherent contribution of the random array factor is locally suppressed. Consequently, the radiation pattern exhibits a low-sidelobe region around the main beam. Beyond the first zero of $J_1(\cdot)$, the Bessel term changes sign and the suppressed spectral energy gradually reappears, leading to a progressive transition towards the conventional random-array sidelobe floor. The boundary of the low-energy region is determined by the first zero of
$J_1(\cdot)$, i.e.,

\begin{equation}
|\mathbf{q}|_c
=
\frac{j_{1,1}}{r_{\min}},
\qquad
j_{1,1}=3.8317,
\end{equation}
Expressing the spatial frequency in the normalized $uv$ plane as $u
=
\frac{|\mathbf{q}|}{k_0}$ yields


\begin{equation}
u_c
=
\frac{|\mathbf{q}|_c}{k_0}
=
\frac{j_{1,1}}{2\pi}
\frac{\lambda}{r_{\min}}
\approx
0.61
\frac{\lambda}{r_{\min}}.
\label{eq:uc}
\end{equation}
It should be noted that this prediction is derived under the assumption of an ideal hard-core point process, whose pair-correlation function is approximated by a step function. Consequently, its accuracy may decrease for finite arrays or satellite layouts that deviate from this idealized spatial model

Outside this region, the structure factor gradually converges to unity, $
S(\|\mathbf{q}\|)\rightarrow1,$ and (\ref{eq:christian}) naturally reduces to the classical result obtained by Lo~\cite{lo1964},

\begin{equation}
\mathbb{E}\!\left\{|AF_{\mathrm{rand}}|^2\right\}
\simeq
\frac{1}{N_s}.
\label{eq:Lo}
\end{equation}

Classical random-array theory provides analytical expressions for the average sidelobe statistics, whereas the prediction of the peak SLL remains considerably more challenging. Recent statistical approaches, such as \cite{kook2002}, model the peak sidelobe distribution but do not provide a simple closed-form expression directly relating the expected peak SLL to the array physical parameters. Extending the proposed framework towards such a prediction is left for future work.

The impact of the swarm radiation pattern on the user SINR, and consequently on the system sum rate, strongly depends on the spatial distribution of the users. Deriving a general analytical expression would require restrictive assumptions on the user distribution and scheduling strategy, limiting its applicability to particular deployment scenarios. For this reason, the performance evaluation presented in this work relies on Monte Carlo simulations. Nevertheless, the proposed analytical model provides useful qualitative insights into the expected system behaviour. For sparse user distributions, e.g., users uniformly distributed over a large coverage area, the probability that two simultaneously scheduled users fall within each other's main beam is low for both collocated and larger distributed swarms. As a result, inter-user interference is mainly determined by the SLL and may remain relatively weak compared to the thermal noise, so  only modest differences in SINR and sum rate are expected between the different array configurations. In contrast, for dense user hotspots  interference becomes quickly dominated by the main lobe. In this regime, the narrower beams provided by larger distributed apertures can spatially separate users more effectively, while the low sidelobe region predicted by the proposed analytical model further limits the interference between nearby beams. Therefore, distributed swarms are expected to provide substantial SINR and sum-rate improvements over conventional collocated arrays. Practical deployment scenarios generally lie between these two extreme cases. Consequently, the achievable performance is ultimately determined by the trade-off between beamwidth reduction and sidelobe behaviour, which is numerically evaluated in Section \ref{secVII}. 
\color{black}

\subsection{Evaluated swarm configurations}
\begin{figure*}[t]
    \centering
    \includegraphics[width=0.7\textwidth]{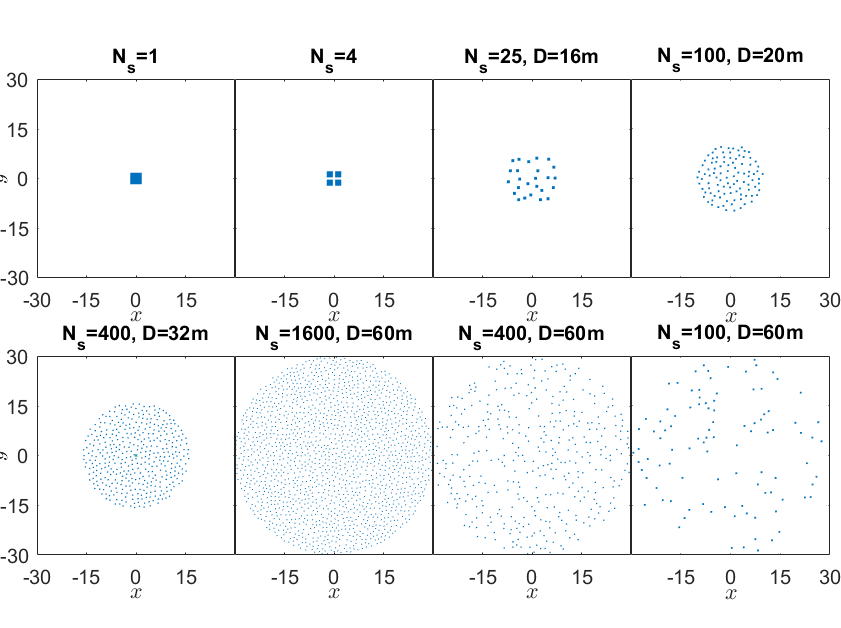}
    \caption{Array configurations with $N=1600$ radiating elements. Axes are expressed in meters. Diameter of random swarms is set according to $D_{min}$ except for the last two cases, which use larger ISS}
    \label{fig:arrays}
\end{figure*}

We consider arrays with $N$ radiating elements, organized in $N_s\leq{N}$ satellite nodes, each equipped with a  uniform square array of $N_\mathrm{sub}=N/N_s$ antennas  with an inter-element separation close to half the wavelength, so grating lobes at satellite node level fall out of the coverage area of the satellite system. The radiation pattern for each radiating element is assumed to follow a $cos^2(\theta_k)$ function. Three main configurations are considered:
\begin{itemize}
    \item $N_s=1$, which corresponds to the monolithic aperture
    \item $N_s=4$, which approaches the FoA concept of \cite{gaud25}. In this case, a uniform square arrangement of satellites is assumed with inter-satellite  spacing between satellite edges equal to $ISS_{\mathrm{min}}$
    \item $N_s\gg4$, which extends the FoA concept for satellites with smaller arrays down to the single-antenna case. Here we consider that the satellites are randomly distributed within a circle of diameter \textit{D} keeping a minimum inter-satellite separation between satellite edges of  $ISS\geq{ISS_{\mathrm{min}}}$ \cite{art25}.
\end{itemize}

The random satellite position for the later case can be calculated following an iterative process. In each step, a new position candidate is randomly generated, the distances to all the already existing nodes in the swarm are calculated, and the candidate is accepted if the minimum distance is above $ISS_{\mathrm{min}}$. In practice, any random distribution fulfilling the minimum distance criterion can be targeted during the initial swarm deployment, whereas during the formation flying, the nodes do not need to keep that exact position, but just a minimum distance above $ISS_{\mathrm{min}}$ to their neighbours and a maximum distance to the overall array centre below $D/2$. This may simplify in-orbit operations. Remarkably, for a given $N_s$ and $ISS_{min}$, there is a minimum feasible swarm diameter $D_{\mathrm{\mathrm{min}}}(N_s,ISS_{\mathrm{min}})$. Indeed, the maximum packing density of equal circular footprints in a plane is achieved by a hexagonal lattice, yielding a packing efficiency of $\pi/(2\sqrt3)=0.9$ \cite{fej64}. To provide sufficient room for random satellite placement and formation-flying corrections, lower efficiencies need to be considered. Moreover,  due to boundary effects, the achievable packing efficiency degrades as the number of satellites decreases. The aforementioned iterative process is used to empirically estimate practical packing efficiencies, which are found to be between two and nine times lower than the theoretical hexagonal packing efficiency, hence

\begin{figure*}[!t]
    \centering
    \includegraphics[width=0.8\textwidth]{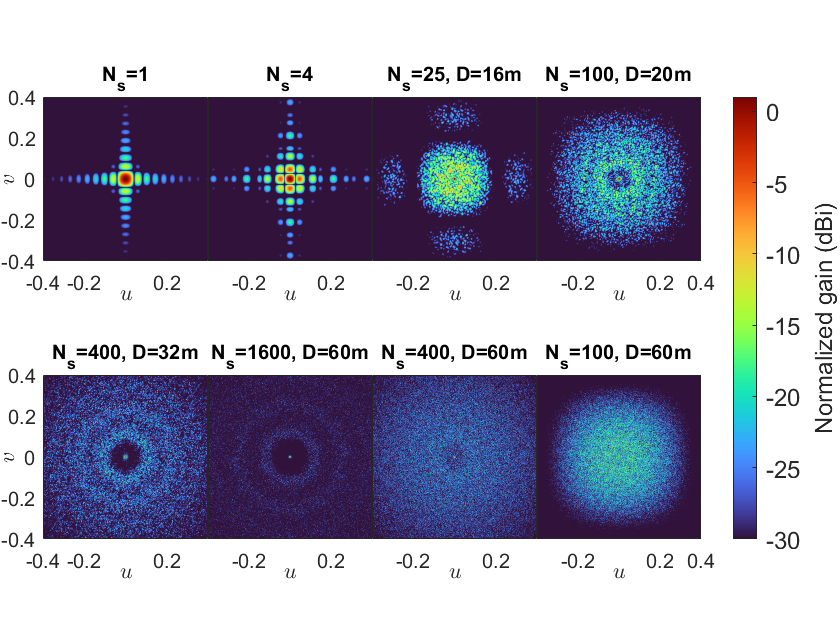}
    \caption{Radiation patterns for different array array configurations with $N=1600$ radiating elements}
    \label{fig:patterns}
\end{figure*}

\begin{equation}
0.1\leq\frac{N_s\pi ISS_{\mathrm{min}}^2}{\pi D_{\mathrm{min}}^2}\leq0.45,
\label{eq20}\end{equation}
 and solving for $D_{\mathrm{min}}$
 
 \begin{equation}
ISS_{\mathrm{min}}\sqrt\frac{N_s}{0.45}\leq D_{\mathrm{min}}\leq ISS_{\mathrm{min}}\sqrt\frac{N_s}{0.1}.
\label{eq20b}\end{equation}

The number of radiating elements $N$ must be set according to the required EIRP per user. Indeed, it is assumed that each radiating element is fed by a solid-state amplifier with transmit power $P_e$, and hence, the total EIRP is $\mathrm{EIRP}=N^2P_eG_n$. Note that here we are also implicitly assuming that the available power per satellite scales with the number of elements, which is aligned with the discussion in \cite{bac23} that considers an available power proportional to the array area. 

\textcolor{black}{To evaluate the trade-off between beamwidth, SLL and sidelobe spatial distribution}, Fig. \ref{fig:patterns} shows the broadside radiation patterns of the array configurations of Fig. \ref{fig:arrays}, where the minimum distance between satellites is set to $ISS_{\mathrm{min}}=1$ m. Whereas swarm diameter has been set to estimated $D_{min}$ value  for the first six configurations, it has been increased to 60 m in the last two cases, keeping a reduced number of satellites, to assess the impact of using larger ISS values. Table \ref{tab:array} summarizes the 8 configurations under consideration including the resulting half-power beamwidth (HPBW) and the footprint radius (FPR) at nadir. 

Following \eqref{eq:HPBWdeg}, the beamwidth get reduced as the aperture increases, enabling more aggressive spatial reuse.  Distinct grating lobes are only observed for $N_s$ = 4, whereas in the other swarm configurations, the randomized geometry spreads the grating-lobe energy across the sidelobe region determined by the radiation pattern of the subarrays in each satellite node. Increasing the subarray size results in higher sidelobe levels, owing to both the increased subarray gain and the reduction of $N_s$
 in \eqref{eq:Lo}. However, these higher sidelobes are concentrated over a narrower angular region, \textcolor{black}{illustrating the trade-off between SLL and spatial localization}.
 
\textcolor{black}{When the swarm diameter is set to the minimum feasible value $D_{\mathrm{min}}$, the low-sidelobe region predicted by \eqref{eq:Sk} can be clearly identified around the main beam. For $ISS_{\mathrm{min}}=1$~m, which corresponds to $r_{\mathrm{min}}=1$ m in \eqref{eq:uc}, and an operating frequency of 2.2~GHz, \eqref{eq:uc} predicts a normalized radius of approximately $0.083$, which agrees well with the results obtained for the $D=60$~m, $N_s=1600$ and $D=32$~m, $N_s=400$ configurations. For smaller swarms, however, boundary effects reduce the achievable packing efficiency, resulting in a larger effective exclusion radius than the nominal $ISS_{\mathrm{min}}$, leading to a smaller low-sidelobe region than predicted by the ideal hard-core model.
For the last two configurations with $D=60$~m and a reduced number of satellites, the lower spatial density weakens the influence of the minimum-distance constraint, making the satellite distribution increasingly resemble an i.i.d.\ uniform process. As a result, the low-SLL region predicted by \eqref{eq:Sk} progressively disappears.}

Remarkably, the single-antenna configuration provides both the lowest sidelobe levels and the widest low-sidelobe region around the main beam.

\begin{table}[ht]
\caption{Array configurations with 1600 ratiating elements}
\centering
\begin{tabular}{cccc}
\hline
\textbf{$N_s$}& \textbf{$D (m)$}& \textbf{HPBW}& \textbf{FPR (km)} \\
\hline
 1 & 4.6 &2.12$^\circ$& 11.1 \\
 4 & 6 &1.44$^\circ$& 7.5\\
 25 & 16 &0.52$^\circ$& 2.7 \\
100 & 20 &0.4$^\circ$& 2.1\\
 400 & 32 &0.22$^\circ$& 1.2 \\
 1600 & 60 &0.12$^\circ$& 0.6 \\
 400 & 60 &0.12$^\circ$& 0.6 \\
 100 & 60 &0.12$^\circ$& 0.6 \\

\hline
\end{tabular}
\label{tab:array}
\end{table}
\section{Beamforming design}
\label{secIV}
In large scale antenna systems, acquiring CSI becomes a fundamental issue since, in contrasts to terrestrial systems, reciprocity cannot be applied due to FDD operation. Moreover, closed loop mechanisms where the satellite system transmits orthogonal reference signals from each antenna or beam in a grid \cite{art24} are not feasible due to the large number of antennas or beams required. Therefore, we assume a location-based scheme \cite{gui24}, \cite{wan26}, in which the users periodically report their position to the satellite system. We assume that users report their initial position immediately after completing the random access procedure. It should be noted, however, that random access in distributed satellite swarms is particularly challenging due to the narrow beamwidth  relative to the overall field of view of the swarm \cite{tuz23}. The design and analysis of random access mechanisms for large-scale satellite swarms are beyond the scope of this paper. Therefore, we assume that the central node responsible for the feeder link is also capable of managing the random access process.  
Two basic beamforming schemes are considered. The first is the MRT\footnote{Strictly speaking, MRT employs the conjugate of the channel vector as the beamforming weight. In this work, however, only the conjugate of the LOS phase component is used with equal power allocation per user, since instantaneous amplitude variations are not available and the Rician K-factor is high. This scheme is denoted as the conventional beamformer in \cite{gui24}. Nevertheless, we retain the term MRT to facilitate comparison with related works such as \cite{gaud25} and \cite{tuz23b}.}, whose weights for the $q$-th subcarreir are calculated as
\begin{equation}
\mathbf{b}_{kn,q}=\sqrt{\frac{P_e}{N_{sc}}}e^{j2\pi\frac{ f_q}{c} \left( \hat{d}_n \cos \hat{\varphi}_n \hat{u}_k+\hat{d}_n \sin \hat{\varphi}_n \hat{v}_k\right)}.
\label{eq24}\end{equation}
The hat symbol over the variables controlling the position of users and antennas denote that they are estimations that can differ from real values, as discussed in Section \ref{secVI}.

The second is the MMSE beamformer. Its unnormalized matrix expression is

\begin{equation}
\color{black}
\mathbf{\bar{B}}=\left(\mathbf{\hat{A}}^H\mathbf{\hat{A}}\frac{P_eN}{KN_{sc}}+\mathbf{I}_N\right)^{-1}\mathbf{\hat{A}}^{\mathrm{H}}
\label{eq25}\end{equation}
where $\mathbf{\hat{A}}=\left(\mathbf{\hat{a}}_{1,LOS},...,\mathbf{\hat{a}}_{k,LOS}\right)^T $ denotes the estimated $\mathbb{C}^{K\times N}$channel matrix that accounts for the deterministic magnitudes and phases associated with the user–satellite geometry. Its elements can be calculated as in \eqref{eq:An} and \eqref{eq:phase}, but again considering that user and antenna positions are estimated. To fulfil the per antenna power constraints, we resort to a double normalization by rows and columns, which exhibits superior performance to alternative approaches under the low-SNR conditions commonly encountered in satellite downlink transmissions \cite{bac23,ang20}. Therefore,

\begin{equation}
\mathbf{{B}}=\mathbf{D_r}\mathbf{\bar{B}}\mathbf{D_c}
\label{eq26}\end{equation}

with 

\begin{equation}
\mathbf{D_c} =
\left(
\mathrm{diag}
\left[
\sum_{n=1}^{N} \left| \bar{b}_{n,1}\right|^2,
\;\cdots,\;
\sum_{n=1}^{N_R} \left| \bar{b}_{n,K}\right|^2,
\right]
\right)^{-1/2}
\end{equation}

\begin{equation}
\mathbf{D_r} = \sqrt{P_e
}\left(
\mathrm{diag}
\left[
\sum_{k=1}^{K} \left| \bar{b}_{1,k}\right|^2,
\;\cdots,\;
\sum_{k=1}^{K} \left| \bar{b}_{n,K}\right|^2,
\right]
\right)^{-1/2}
\end{equation}
In the latter two expressions, the subcarrier index has been omitted for the sake of readability.

\section{User distribution and scheduling}
\label{secV}
The performance achieved by different distributed swarm configurations depends not only on their radiation patterns, as discussed in Section \ref{secIII}, but also on the distribution of the simultaneously served users. Subsections below present the considered distributions and scheduling algorithms.
\subsection{User distribution}
The following three user distribution scenarios are considered. 
\begin{itemize}
    \item uniform: $K$ users are uniformly distributed within the region covered by $\lvert  \theta \rvert\leq\theta_{\text{max}}$. The distribution is done in geodetic coordinates, i.e. latitude and longitude, which yields to larger user densities at coverage edges. 
    \item single hotspot: $K$ users are uniformly distributed (in geodetic coordinates) within a region covered by $\lvert\theta_k\rvert\leq\theta_{hs_{\text{max}}}$ around a central position  $(\theta_{hs_c},\phi_{hs_c}) $ with $\theta_{hs_c}\leq\theta_{\text{max}}$. 
    \item clustered: $\lfloor Kp_{hs} \rfloor$ users are evenly split in $N_{hs}$ identical hotspots with centres $(\theta_{hs_c}(n_{hs}),\phi_{hs_c}(n_{hs}))$  uniformly distributed within the coverage area ($\lvert\theta_{hs_c}\rvert\leq\theta_{\text{max}}$). Within each hotspot, users are uniformly distributed in the region $\lvert\theta_k\rvert\leq\theta_{hs_{\text{max}}}$. The remaining 1-$\lfloor Kp_{hs} \rfloor$  users are uniformly distributed within the remaining coverage area. Two examples of this distribution are shown in Fig. \ref{fig10}.
\end{itemize}
User height is set to sea level in all cases. The two first scenarios are more academic and are used to identify performance trends. The later is a tractable simplification of more realistic clustered scenarios \cite{fil24}.

\subsection{User scheduling}
Regarding user scheduling, two approaches are considered. In the first, no scheduling is performed, i.e. all users in the coverage area are simultaneously served using all time-frequency resources leading to $f_{\mathrm{sch},k}=1\, \forall\, k$. In the second, user scheduling is used to keep outage probability arbitrarily low. We resort to a Time Domain Multiple Access (TDMA) scheme in which different groups of users are served in different time slots. User grouping is performed following a minimum distance criterion, which forces users with angular separations below the system beamwidth to be scheduled in different groups. Note here that the minimum distance threshold depends on the swarm configuration. In contrast to \cite{bac23}, we consider geometrical angular distance and a TDMA scheme that allows serving the same user in multiple slots as far as it fulfils the minimum distance criterion with all the other group members.  A maximum number of users per group $K_{g_{\max}}$ is set to avoid increasing outage probability due to low SNR conditions, since the total available power is shared by all users. The proposed algorithm is described in Algorithm 1.

\begin{algorithm}
\caption{Minimum distance scheduling}

\begin{algorithmic}[1]
\STATE \textbf{Input:} Vector of angular coordinates $\theta, \phi$; threshold $d_{th}$ (minimum distance in u-v plane); maximum group size $K_{g_{\max}}$
\STATE Compute projection to u-v plane
\STATE Compute pairwise euclidean distance matrix $d(i,j)$
\STATE Sort distances and identify pairs closer than $d_{th}$
$\mathcal{W} = \{(i,j): d(i,j) < d_{th}\}$.  Users not included in any pair are grouped in \[
\mathcal{U}_{\mathrm{rem}} =
\{ k: \forall (i,j) \in \mathcal{W}, \ k \neq i , k \neq j \}
\]
 \STATE Initialize two groups clusters, each with a different user of the closest pair
 \STATE \textbf{Assign remaining elements in $\mathcal{W}$:}
    \FOR{each unassigned element $i$}
        \FOR{each group $g \in \mathcal{G}$}
            \IF{$\min_{j \in \mathcal{G}} d_{i,j} \geq d_{th} \quad \& \quad |g|<K_{g_{\max}}$}
                \STATE Assign $i$ to group $g$ 
                \STATE Break
            \ENDIF
        \ENDFOR
        \STATE If not assigned, create new group
    \ENDFOR
    \STATE \textbf{Equalize group sizes}:
        \REPEAT
        \STATE Compute group sizes
        \STATE Move elements between largest and smallest groups if minimum distance constraint satisfied
    \UNTIL{no changes possible or maximum difference between groups of 1 user}
    
\STATE \textbf{Assign remaining users not included in $\mathcal{W}$ to all groups:}
\FOR{each group $g \in \mathcal{G}$}

    \IF{$|g| + |\mathcal{U}_{rem}| \leq K_{g_{\max}}$}

        \STATE Assign all users in $\mathcal{U}_{rem}$ to group $g$

    \ELSE

        \STATE Compute number of available positions:
        $
        N_{free} = K_{g_{\max}} - |g|
        $

        \STATE Select the next $N_{free}$ users from $\mathcal{U}_{rem}$

        \STATE Assign selected users to group $g$

        \STATE Update allocation index for remaining users

    \ENDIF

\ENDFOR
\STATE \textbf{Output:} groups $\mathcal{G}$

\end{algorithmic}
\end{algorithm}

It is worth mentioning that in a real system, user scheduling depends on user demands. Here we assume the same  demand among all users, which allows us comparing the impact of different swarm configurations. \textcolor{black}{Therefore, all groups are served sequentially in a round-robin fashion, and the scheduling factor in \eqref{eq21} corresponds to the ratio of the number of groups in which a user is scheduled to the total number of groups.}
\section{Positioning and synchronization errors}
\label{secVI}
Large distributed apertures face three important implementation challenges that affect their beamforming performance:(i) synchronization among distributed nodes; which is far more difficult than within a monolithic aperture; (ii) acquiring a precise knowledge of the relative position of the distributed antennas, which is not present in monolithic apertures since the positions are physically fixed; and (iii) acquiring precise knowledge of the position of users, which is shared with monolithic apertures, but it can have larger impact due to the narrower bandwidths in distributed approaches. In practice, they lead to position and synchronization errors that, as anticipated in Section \ref{secIV}, translate to differences between the real channel matrix and the estimated one used for calculating the beamformers. In presence of such errors, the coefficients of the estimated matrix $\mathbf{\hat{A}}$ can be modelled as
\begin{equation}\begin{array}{rl}
\hat{a}_{n} \left(\theta_k,\phi_k \right)=&\sqrt{\displaystyle\frac{G_n\left(\hat{\theta_k},\hat{\phi_k} \right) G_R}{\text{PL}(\hat{\theta_k} ) \text{OL}}} \times \\
&\displaystyle e^{-j2\pi \frac{f_c}{c} \left( \hat{d_n} \sin \hat{\theta_{k}} \cos \left(\hat{\phi_k}-\hat{\varphi_n} \right)\right)} \times \\
&\displaystyle e^{-j\frac{\pi}{180}\epsilon_s};
\end{array}\label{eq29}\end{equation}
\textcolor{black}{where $\hat{\theta}_k=\theta_k+\epsilon_{\theta_k}$ and
$\hat{\phi}_k=\phi_k+\epsilon_{\phi_k}$. The perturbed position of the
$n$-th antenna is modelled as
\begin{equation}
\hat{\mathbf{p}}_n=
\mathbf{p}_n+
\epsilon_{d_n}
\begin{bmatrix}
\cos\psi_n\\
\sin\psi_n\\
0
\end{bmatrix},
\end{equation}
where $\epsilon_{d_n}\sim\mathcal{N}(0,\sigma_d^2)$ denotes the signed
displacement amplitude and $\psi_n\sim U[0,\pi)$ its direction. The
perturbed polar coordinates $(\hat{d}_n,\hat{\phi}_n)$ are obtained from
$\hat{\mathbf{p}}_n$ using \eqref{eq1}.The adopted model considers position uncertainties within the swarm plane, consistently with the planar distributed aperture assumed throughout the manuscript. Extending the model to three-dimensional position errors is straightforward, as the resulting phase perturbation depends on the projection of the position error onto the propagation direction. Such an extension is left for future work.}

Imperfect satellite synchronization is modelled as a phase error \cite{gaud25} denoted by $\epsilon_s$, which follows a normal distribution $\mathcal{N}(0,\sigma_s^2)$. The different error sources are assumed to be independent. Let us remark that the errors are exactly the same for all antennas collocated in a given satellite node, so satellite attitude or synchronization errors within each monolithic array are not considered.

 The errors on the relative angular position of users $\epsilon_{\theta k}$ and $\epsilon_{\varphi k}$ depend on three factors: (i) the accuracy of the user positioning, which is assumed to be well below the half-power beamwidth and is thus neglected; (ii) the rate at which the users provide position updates; and (iii) the rate at which the satellite updates the beamforming coefficients. The impact of the latter two is analysed in Section \ref{secVII}.
\section{Numerical evaluation}
\label{secVII}
The scenario under evaluation considers a LEO satellite orbiting at 600 km and serving users with a minim elevation angle of 40$^\circ$ at S-Band and with an operating bandwidth of 20 MHz. Table \ref{tab:params} summarizes the used simulation parameters, unless otherwise stated.
Available literature offers very different transmit power per element values. Reference \cite{bac23} considers almost -12 dBW for a large formation of arrays scheme, though the same authors increase it to -8.3 dBW in \cite{gaud25}. In contrast, \cite{tuz23b} considers -4.6 dBW obtained from current cubesat technology. Here we assume this latter value.
The number of elements has been chosen by considering a target EIRP per beam of 48 dBW \cite{nr821}. Using 3GPP recommended layout \cite{nr821}, 379 beams are needed to fill the target coverage area, and assuming that 10$\%$ can be simultaneously illuminated leads to 38 active beams, yielding a total target EIRP of 64.6 dBW. With $P_e$=-4.6 dBW, 1530 elements are needed to achieve the total target EIRP, which are rounded to 1600 to simplify the formation of square arrays. 
The minimum inter-satellite spacing has also been assumed differently in the literature, from 100 m in \cite{tam26} to around 1 m in \cite{gaud25}, \cite{tuz23b}, \cite{had16}. In-orbit flying formation experiments such as \cite{bod12}, demonstrated high accuracy positioning with inter-satellite distances as low as 2 m or 3 m. Here we assume $ISS_{\mathrm{min}}$=1 m for the direct comparison to most relevant prior works, though the impact of larger inter-satellite distances is also discussed by the last two array configurations in Table \ref{tab:array}.
The OL value obeys to 2.2 dB of scintillation loss \cite{nr821} and an extra factor of 4 dB that accounts for RF losses, demodulator losses or polarization losses \cite{gaud25}, considering that the two antenna scheme of \cite{nr821} will still present combining losses. The minimum SINR threshold required to avoid outage is set to $\mathrm{SINR}_{\mathrm{th}} = -7\,\text{dB}$ \cite{gaud25}.

\begin{table}[ht]
\caption{System parameters}
\centering
\begin{tabular}{cc}
\hline
\textbf{Parameter} & \textbf{Value} \\
\hline
Orbital altitude & 600 km \\
Max scan angle ($ \theta_{max}$) & 44.4$^\circ$ \\
Central frequency ($f_c$) & 2.2 GHz\\
Bandwidth ($B_w$) & 20 MHz\\
Number of radiating elements ($N$) & 1600\\
Tx power per element ($P_e$) & -4.6 dBW\\
Element gain & 5.5 dBi\\
Element spacing (same satellite) & 0.6$\lambda$\\
Minimum inter-satellite spacing ($ISS_{\mathrm{min}}$) & 1 m\\
User G/T & -31.6 dB/K\\
Rician K-factor & 10 dB\\
Other loss ($OL$) & 6.2 dB\\
SINR outage threshold ($\mathrm{SINR}_{\mathrm{th}}$)& -7 dB\\
OFDM subcarriers ($N_{sc}$) & 40\\
\hline
\end{tabular}
\label{tab:params}
\end{table}

The numerical analysis is performed for the array configurations summarized in Table \ref{tab:array} over four different scenarios regarding user distribution, whose parameters are detailed in Table \ref{tab:scens}.

\begin{table}[ht]
\caption{User distribution scenarios}
    \centering 
   \small
\begin{tabular}{p{2cm}>{\centering}p{0.75cm}>{\centering}p{0.75cm}>{\centering}p{0.75cm}p{0.75cm}}
\hline
\textbf{Parameter} & \textbf{Uniform} & \textbf{Single-hotspot}  & \textbf{Clustered 1}  & \textbf{Clustered 2}\\
\hline
Number of hotspots ($N_{hs}$) & 0 & 1 & 16 & 4 \\
Hotspot size ($\theta_{hs_{max}}$) & - & 10$^\circ$ & 0.625$^\circ$ & 2.5$^\circ$ \\
Hotspot radius at $\theta_k=0^\circ$ (km) & - & 106  & 6.6  & 26.2  \\
Percentage of users in hotspots ($p_{hs}$) & 0 & 100 & 80 & 80\\

\hline
\end{tabular}
\label{tab:scens}
\end{table}

\subsection{Baseline results}
In this section, we compare the different antenna formations in Fig. \ref{fig:arrays} for different user distribution scenarios, assuming no scheduling and perfect synchronization and positioning. \textcolor{black}{The goal is to assess the impact of the radiation pattern trade-offs, including beamwidth and sidelobe levels and distribution, on the communication performance under idealized conditions.}

\begin{figure}
    \centering
    \includegraphics[width=0.45\textwidth]{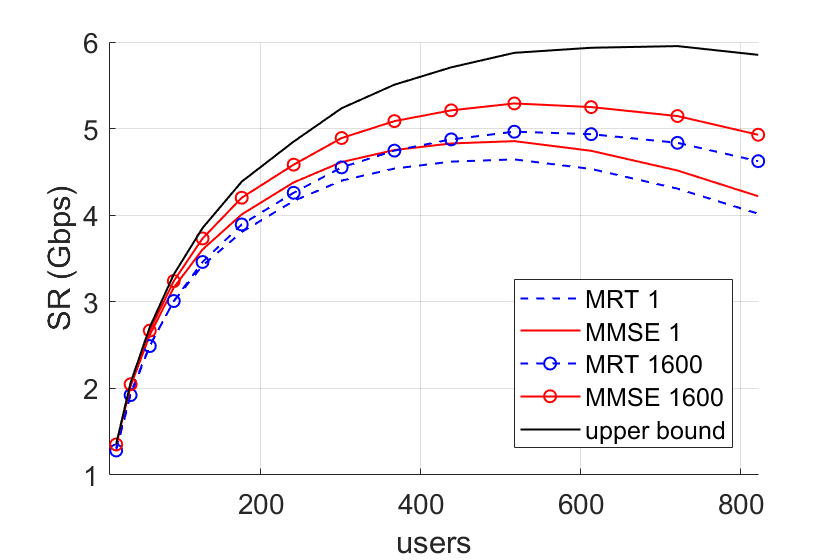}
    \caption{Sum Rate as a function of number of users for uniform user distribution and for two different number of satellite nodes (1 or 1600)}
    \label{fig4}
\end{figure}

\begin{figure}
    \centering
    \includegraphics[width=0.45\textwidth]{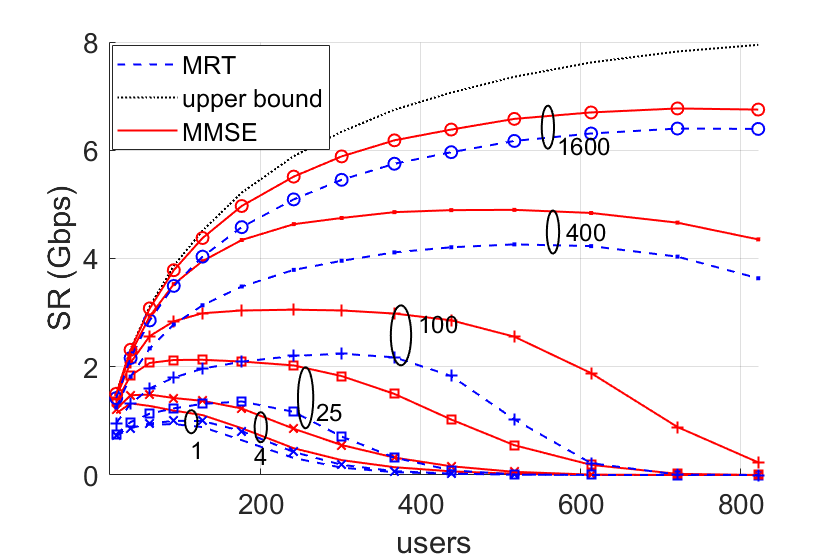}
    \caption{Sum Rate as a function of number of users for a single hotspot and different number of satellite nodes (1 to 1600)}
    \label{fig5}
\end{figure}

Figs. \ref{fig4}-\ref{fig5} show the results of  Monte Carlo simulations for uniform user distribution and a single hotspot of size $\theta_{hs_{max}}=10^\circ$ lying between the boresight and the edge of coverage, at an intermediate off-axis angle  $(\theta_{hs_c},\phi_{hs_c})=(26.9^\circ, 0^\circ)$, respectively. The considered swarm diameters (except for the $N_s=4$ case) are set according to $D_{\mathrm{min}}$ in \eqref{eq20b}. The number of users is continuously increased to identify optimal operating points. Both figures compare MRT and MMSE beamformers with an upper bound (UB) obtained by using the MRT but artificially neglecting any interference. For a uniform user distribution, as predicted, the user density is insufficient to fully exploit the reduced beamwidth of large distributed swarms. The improvement over a monolithic array is not significant, and both array configurations perform close to the UB. Fig. \ref{fig4} shows the extreme configurations only, while others lie in between (not shown).  Indeed, the system is primarily noise-limited, which also explains the small performance differences between MMSE and MRT. Nevertheless, an optimal number of simultaneous users can be identified, beyond which the impact of interference becomes more pronounced. The single-hotspot case depicted in Fig. \ref{fig5}  dramatically changes the results. A monotonic increase of SR with the increase of distributed satellite nodes (and distributed aperture) is observed. Here, only the case with single-antenna nodes is able to keep the noise-limited regime performing close to the upper bound. At the number of users maximizing the SR, and using the MMSE precoder, the configurations with 1600 and 400 satellites achieve SR improvements of 413$\%$ and 271$\%$, respectively, compared to the monolithic array.  It is worth noting that, as anticipated in \cite{gaud25}, the configuration with only 4 large distributed arrays performs similarly to the monolithic array.

\begin{figure}
    \centering
    \includegraphics[width=0.45\textwidth]{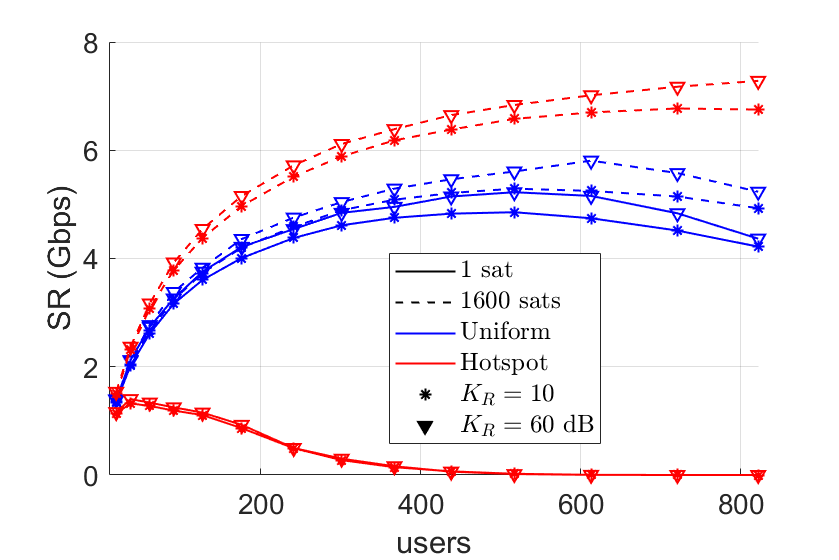}
    \caption{Impact of the NLoS component in the Sum Rate for different swarm configurations and user distributions, and MMSE beamforming.}
    \label{fig:kr}
\end{figure}

\color{black}
Fig. \ref{fig:kr} evaluates the impact of the NLoS channel component. Specifically, it compares two Rician factors, $K_R=10$~dB and $K_R=60$~dB, the latter closely approximating a pure LoS channel. The comparison is carried out for both a uniformly distributed user scenario and a single-hotspot scenario with the same characteristics as in the previous evaluation. Moreover, only the MMSE beamformer and the two extreme swarm configurations, namely a monolithic array and a swarm of 1600 single-antenna satellites, are considered for readability. The results for $K_R=10$~dB exhibit a  degradation with respect to $K_R=60$~dB. Nevertheless, since $K_R=10$~dB still corresponds to a LoS-dominated propagation environment, the resulting performance degradation remains limited. This observation confirms the theoretical discussion in Section \ref{secII}.\ref{secIIb}. As long as the LoS component dominates the propagation channel, the diffuse NLoS component only introduces moderate SINR fluctuations, resulting in a limited impact on both the achievable sum rate and the relative performance comparison among the different swarm configurations.
\color{black}

\begin{figure}
    \centering
    \includegraphics[width=0.45\textwidth]{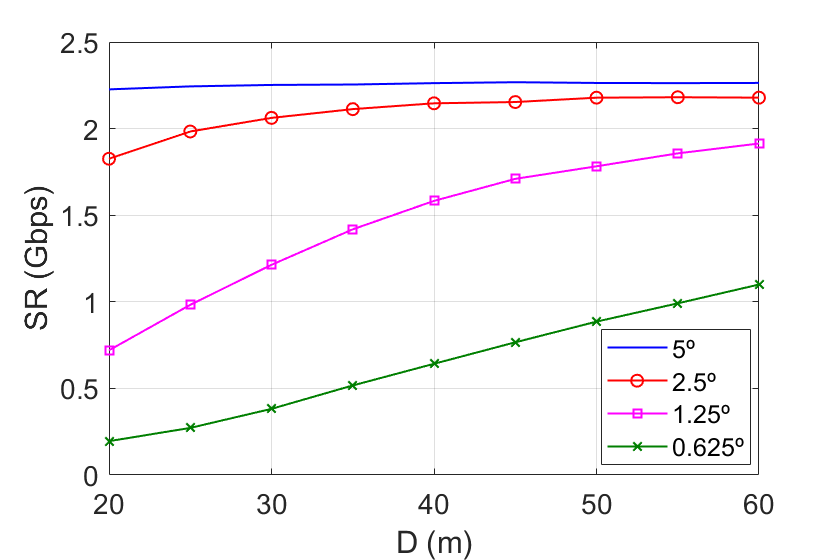}
    \caption{Sum Rate as a function of the aperture diameter for 100 satellites, MMSE beamforming and different hotspots sizes}
    \label{fig6}
\end{figure}

Fig. \ref{fig6} analyses the impact of the size of a single hotspot, at the same location as in the previous analysis, as well as of reducing the satellite density (i.e. increasing ISS) by enlarging the aperture diameter beyond $D_{min}$  while keeping the number of satellites. As a matter of example, we focus on the configuration with 100 satellites serving 50 users. First expected conclusion is that the SR decreases as the hotspot size reduces, since the inter-beam interference grows. The second is that for a given hotspot size, increasing the aperture diameter by increasing ISS provides SR improvements until a saturation point. As the ISS grows, the patterns show smaller beamwidths but higher SLL, and the low-SLL region  close to the  main beam gets reduced (see Fig. \ref{fig:patterns}). Therefore, when this rising of the SLL starts falling in the hotspot, the benefits of the narrow beamwidth get partially compensated \cite{art15}. For instance, for the 2.5$^\circ$ hotspot, increasing $D$ beyond 40 m provides marginal improvement.

\begin{figure}
    \centering
    \includegraphics[width=0.45\textwidth]{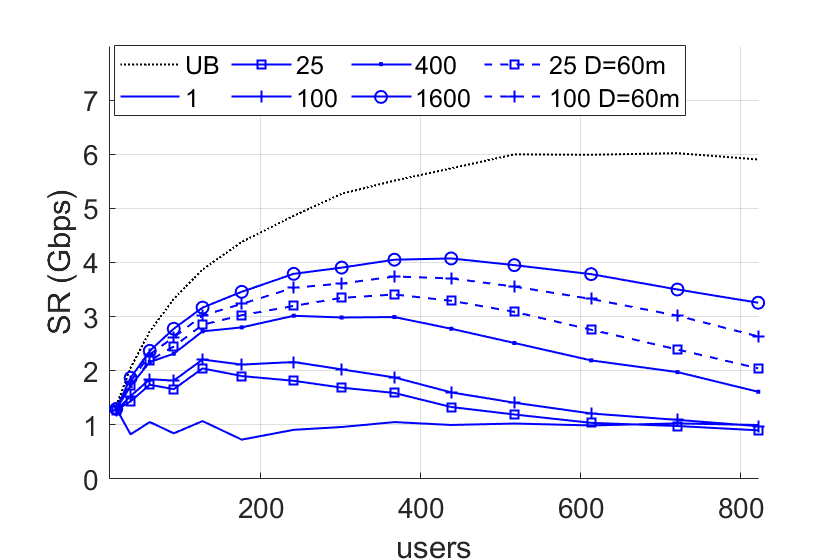}
    \caption{Sum Rate as a function of number of users for clustered scenario 1, different antenna configurations and MRT beamforming }
    \label{fig7}
\end{figure}

\begin{figure}
    \centering
    \includegraphics[width=0.45\textwidth]{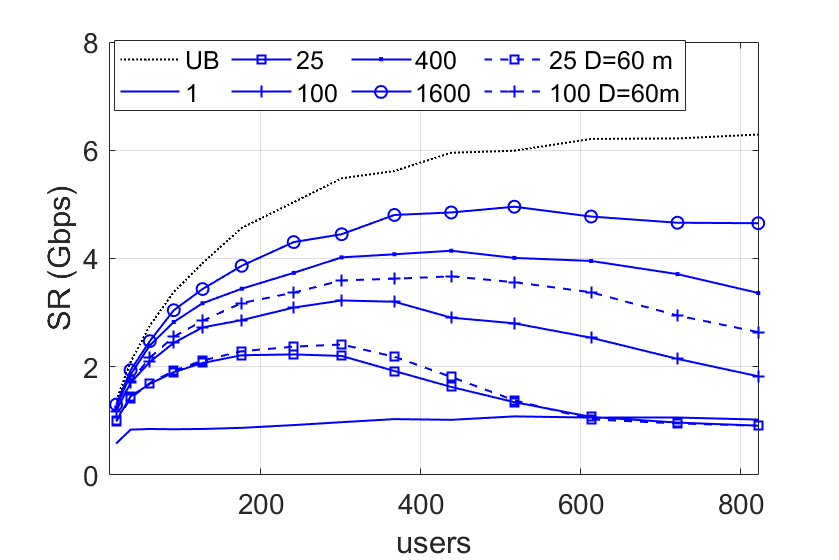}
    \caption{Sum Rate as a function of number of users for clustered scenario 2, different antenna configurations and MRT beamforming}
    \label{fig8}
\end{figure}

From Fig. \ref{fig:patterns}, it is clear that single hotspot scenarios favour distributed swarms since the raise of far SLL does not impact the system performance in terms of increasing the interference. Figs. \ref{fig7}-\ref{fig8} repeat the analysis with clustered scenarios 1 and 2, where hotspots of different sizes coexist with sparsely distributed users. \textcolor{black}{In these scenarios, the interference levels for users within hotspots is primarily determined by the beamwidth, whereas for sparsely distributed users it is mainly governed by the sidelobe distribution. Still,} the monotonic increase of SR with the number of satellites remains in both cases; however, the gap with respect to the upper bound widens, particularly for the clustered-1 distribution. This is attributed to the reduced hotspot size, which increases user density and consequently leads to higher inter-beam interference. This is also the reason why  configurations with 60 m apertures but just 25 or 100 satellites, i.e. with $D>D_{\mathrm{min}}$ present a large SR improvement, even outperforming  the swarm with 400 satellites and $D$=32 m. In contrast, in clustered scenario 2, increasing the aperture diameter beyond $D_{\mathrm{min}}$ provides much smaller gains because the higher sidelobes start overlapping with the hotspot region.

\subsection{Impact of scheduling}
\begin{figure}
    \centering
    \includegraphics[width=0.45\textwidth]{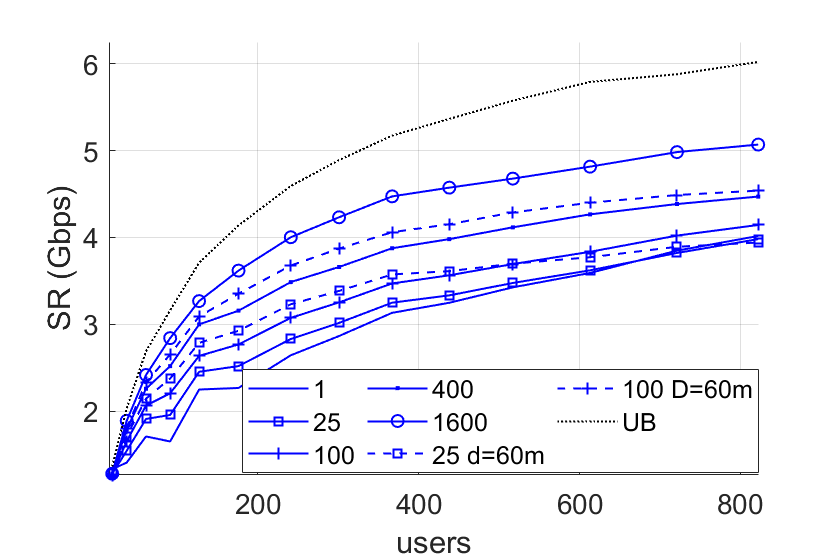}
    \caption{Sum Rate as a function of number of users for different number of array configurations, MRT beamforming, clustered scenario 1 and scheduling algorithm 1}
    \label{fig9}
\end{figure}

\begin{figure}
    \centering
    \includegraphics[width=0.5\textwidth]{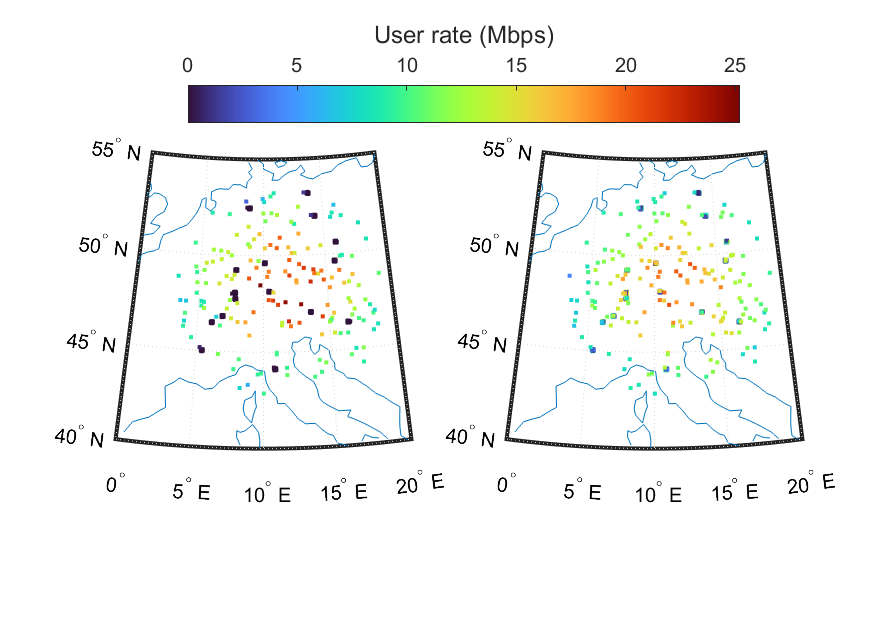}
    \caption{Distribution of user rates for 822 users for MRT beamforming; clustered scenario 1; scheduling algorithm 1; monolithic array (left); and single-antenna swarm with 1600 satellites (right) }
    \label{fig10}
\end{figure}

In this section, we evaluate the minimum-distance scheduling strategy proposed in Algorithm 1, which prevents two users within the same main beam from being scheduled simultaneously, thereby reducing the outage probability. \textcolor{black}{ As a result, the interference within each scheduling slot is mainly governed by the sidelobe distribution. However, a narrower beamwidth generally requires a reduced number of scheduling slots to serve all users, thereby increasing the scheduling factor $f_\mathrm{sch}$.} Fig. \ref{fig9} depicts the results for the more challenging clustered 1 scenario and the MRT. The comparison with Fig. \ref{fig7} shows that the scheduling algorithm provides SR improvements for all configurations, reducing the gap with the upper bound and the differences between monolithic and distributed approaches. \textcolor{black}{Indeed, narrower beams naturally provide a higher spatial separation among users, reducing the dependence of the system performance on the adopted scheduling strategy}. To provide a numerical quantification, at 822 users, the performance gains with respect to no scheduling of the single-antenna swarm and the monolithic array are 9$\%$ and 294$\%$, respectively. However, as demonstrated in Fig. \ref{fig10}, the user rate distribution across the coverage area becomes more uneven as the number of satellites is reduced. Indeed, for the monolithic array, the user rates at the tiny hotspots approach zero, whereas for sparse users located close to the centre of coverage grow up to 25 Mbps. These are the users effectively contributing to the SR improvement. In contrast, the single antenna swarm is able to keep serving the hotspots. Therefore, higher antenna distribution enables a more balanced spatial allocation of capacity across the coverage area.

\subsection{Impact of position and synchronization errors}
\begin{figure}
    \centering
    \includegraphics[width=0.45\textwidth]{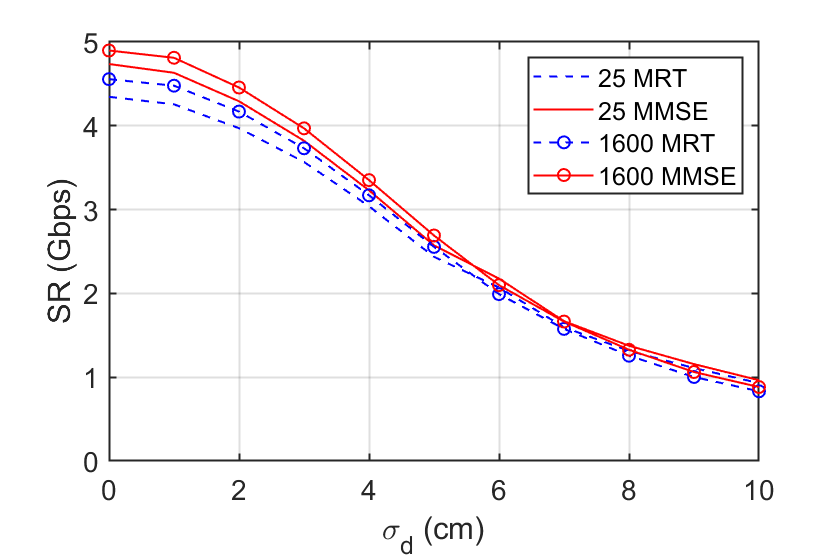}
    \caption{Sum Rate as a function of the error in the estimation of antennas relative position for uniform user distribution and 301 users}
    \label{fig11}
\end{figure}

\begin{figure}
    \centering
    \includegraphics[width=0.45\textwidth]{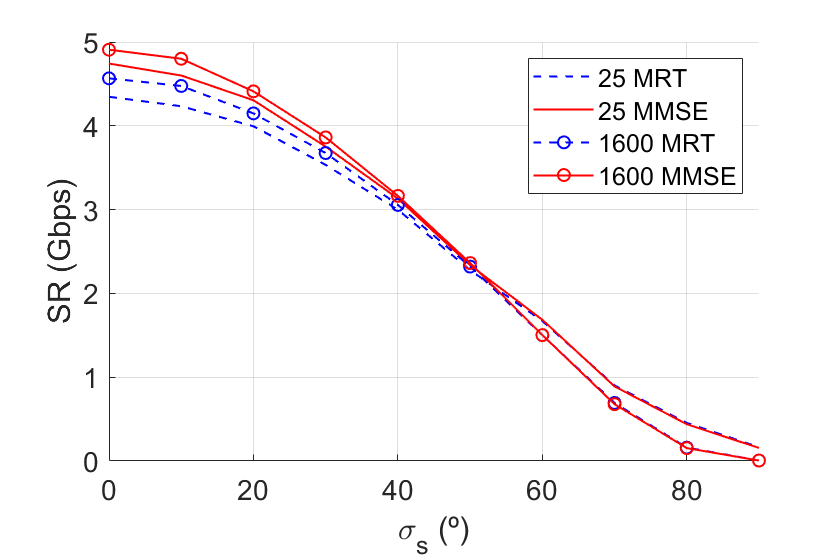}
    \caption{Sum Rate as a function of the synchronization error for uniform user distribution and 301 users}
    \label{fig12}
\end{figure}

This section discusses the conditions required to achieve a significant fraction of the predicted sum-rate improvement under ideal conditions when practical impairments, such as synchronization and positioning errors introduced in Section \ref{secVI}, are taken into account. Figs. \ref{fig11}-\ref{fig12} show the simulated SR as a function of the standard deviation of the error in the relative position of the swarm nodes and on synchronization, respectively. We focus on the smaller and larger random swarm configurations (i.e. $N_s$=25 and $N_s$=1600) and on the uniform user distribution  over the entire coverage area. Remarkably, for the adopted in-plane error model, the antenna position errors  have larger impact as $\lvert\theta_k\lvert$ increases, and at the limit of $\theta_k=0$, they become harmless. Therefore, the uniform distribution is used here to balance the impact of errors across the coverage.

Three main conclusions can be extracted. First, \textcolor{black}{for the considered combinations}, the SR decay does not depend on the beamformer nor on the number of satellite nodes (excluding monolithic arrays). More distribution of satellite nodes and larger apertures does not imply larger error sensitivity. Second, degradation below 10$\%$ can be achieved for phase errors due to synchronization when $\sigma_s\leq20^\circ$. Third, degradation below 10$\%$ requires estimating the position of the satellites with an error fulfilling $\sigma_d\leq2$ cm. This is a hard requirement that cannot be met by standard GNSS receivers. Diverse technologies like optical, ultrawide-band (UWB) or carrier-phase GNSS can achieve  sub-dm precision. Indeed, the later has been in-orbit validated for swarms of two satellites \cite{Mon11}. The scalability of these methods to hundreds or thousands of satellites is still an open research problem, which may ultimately limit the maximum practical distribution of satellites, particularly in the shorter term.

\begin{figure}
    \centering
    \includegraphics[width=0.45\textwidth]{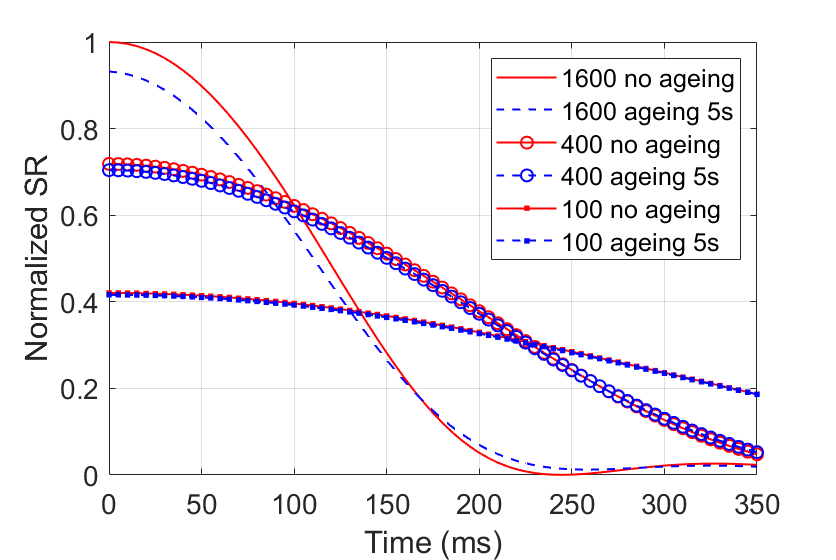}
    \caption{Sum Rate as a function of time for MRT, a single hotspot at  $\theta_{hs_c}=0^\circ$, 300 users, three swarm configurations and two user positioning ageing values.  }
    \label{fig13}
\end{figure}

Finally, we assess the impact of user positioning errors induced by a limited update rate of the user positions and of the beamforming weights. A good GNSS coverage is assumed, so direct position estimation errors are not considered. In particular, we focus on a single hotspot of 10$^\circ$ at $\theta_{hsc}=0^\circ$, which is the worst case location since close to Nadir the beam footprints are smaller and misspointing is more likely to happen. We consider MRT and 300 users with uniformly distributed random speeds between 0 and 350 km/h and directions between 0$^\circ$ and 360$^\circ$. In this case, a pure LoS channel is assumed to avoid the need for complex fading modelling during user movement and to prevent fading effects from masking the impact of mobility.  The coefficients of the channel are then computed setting $K_R=\infty$ in \eqref{eq:rice}.  The other losses $OL$ are increased to 6.7 dB to keep similar SR figures as in previous simulations. We evaluate the SINR degradation as a function of time, assuming that the beamforming weights are updated at t=0 s. Two cases are considered. In the first, the actual user positions at t=0 s are assumed to be instantaneously available and are used to compute the beamforming weights. In the second, the beamformer is computed using user position information reported 5 s earlier, resulting in a 5 s ageing of the position estimates. Fig. \ref{fig13} shows the simulation results for the three largest swarm configurations ($N_s$=1600, $D$=60 m; $N_s$=400, $D$=32 m; $N_s$=100, $D$=20 m). The effects of user-position ageing and limited beamforming update rates are cumulative, although beamforming update requirements are more critical. For the single-antenna swarm with 1600 satellites, a SR degradation below 10$\%$ requires  beamforming update periodicities between 25 ms (user position with 5 s ageing) and 50 ms (no ageing). Moreover, this swarm with 1600 satellites performs better than the next with 400 satellites for update rates below 75 ms. Note, however, that this implies a SR degradation up to 30$\%$, so for such update rate, the 400 satellite solution may be selected  for the sake of stability. Extending this observation, a trade-off emerges: fewer satellites in smaller apertures lead to broader synthesized beams, improving stability and reducing beamforming update requirements, but resulting in lower sum rates. Let us remark that the frame duration in 5G-NTN is 10 ms, so it seems reasonable to consider update rates every one or two frames, hence well supporting the 1600 satellites configuration. 

\begin{figure}
    \centering
    \includegraphics[width=0.45\textwidth]{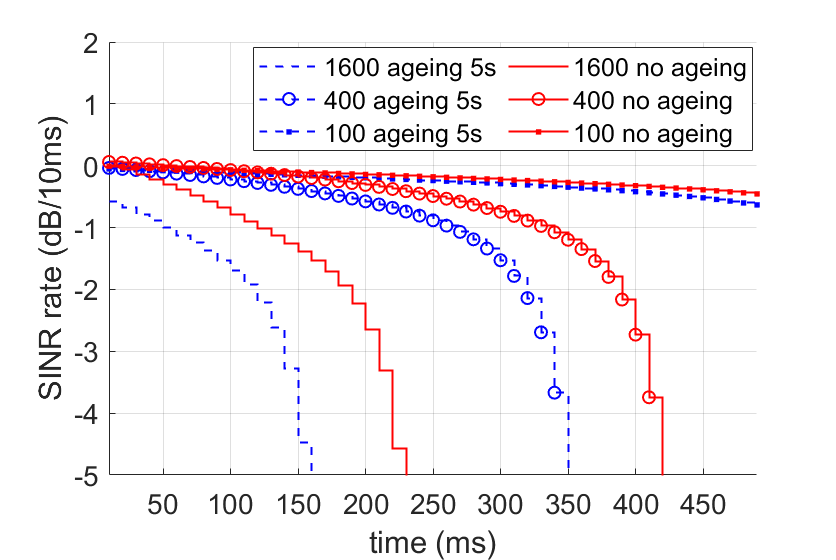}
    \caption{SINR variation in 10 ms window for a single user at $\theta_k=0^\circ$. A single hotspot at $\theta_{hs_c}=0^\circ$; 300 users, three swarm configurations and two user positioning ageing  of 0s and  5 s are considered.  }
    \label{fig14}
\end{figure}

Previous discussion considered stability in terms of sum rate, but at user level, SINR variations require Modulation and Coding  scheme (MCS) adaptation. To assess this, we focus on the worst case scenario, with a target user located at $\theta_k=0^\circ$, moving  at 350 km/h in the direction opposite to the satellite.  The analysis considers the same single-hotspot scenario with 300 users, MRT beamforming, and user position information subject to either no ageing or 5 s ageing. Fig. \ref{fig14} shows the variation of the target user SINR over 10 ms windows as a function of time. The SINR degradation rate increases as the user approaches the first null of the swarm radiation pattern. Beyond the first null, the SINR rate starts to increase; however, this region is omitted for clarity. To interpret these results, we assume that the MCS is updated once per frame, i.e., every 10 ms, and that SINR variations of up to 1 dB are tolerable within a given MCS. Larger SINR fluctuations over a 10 ms interval may significantly impact the Block Error Rate (BLER). Under this assumption, beamforming updates are required every 50 ms for the 5 s ageing case and every 120 ms when no ageing is present in a single-antenna swarm comprising 1600 satellites. Larger MCS adaptation periodicities would require faster beamforming updates or selecting more stable array configurations like 400 satellites. Again, reducing the number of satellites and hence, the resulting aperture, provides more robust operation against mobility, but de 60 m swarm is still well supported.

\section{Conclusions}
\label{secVIII}
This work analysed distributed satellite swarms for D2C applications as an alternative to conventional monolithic arrays, with the objective of fully exploiting the higher spatial resolution enabled by large distributed apertures. Different swarm configurations and monolithic arrays with the same number of antenna elements and identical EIRP were compared. \textcolor{black}{ For the considered scenarios, the results showed that the most distributed architecture, consisting of single-antenna satellites, consistently outperformed all other configurations under ideal operating conditions, i.e. assuming perfect synchronization and satellite-user positioning.}  While the performance improvement over monolithic arrays was moderate under uniformly distributed user conditions, the advantages of distributed apertures became significantly more pronounced as users were concentrated in hotspot areas. The limited spatial resolution of smaller apertures could be partially mitigated through minimum-distance scheduling strategies, reducing the sum-rate gap with respect to larger distributed apertures. Nevertheless, such approaches remained insufficient to efficiently serve users in dense hotspots, mainly benefiting sparsely distributed users instead. Likewise, increasing the swarm aperture solely by enlarging the inter-satellite spacing while maintaining a small number of satellite nodes yielded gains in scenarios with very narrow hotspots. However, the resulting increase in sidelobe levels prevented achieving the performance attainable with swarms of comparable aperture but higher node density.

Despite the observed performance advantages, large distributed swarms revealed several important technical challenges. First, maintaining coherent beamforming required update periodicities on the order of a few tens of milliseconds. Although demanding, such requirements were compatible with current 5G frame durations of 10~ms. Longer update intervals would require reducing the effective swarm aperture, thereby improving link stability at the expense of capacity. Second, centimetre-level relative positioning accuracy within the swarm was required. While this requirement was largely independent of the swarm configuration, scaling precise positioning and ranging solutions to thousands of distributed nodes remains a major open research challenge. Similarly, maintaining phase synchronization errors below $20^\circ$ presents significant scalability issues for large-scale distributed arrays. Both issues may ultimately limit the maximum achievable swarm aperture. Other system-level aspects also remain open challenges and require further investigation. These include the random access procedure over wide fields of view, which is complicated by operation under very narrow bandwidths, and the efficient distribution of data from the centralized node to the swarm.

Although distributed swarms composed of hundreds or thousands of satellites are still impractical with current technology, they should be regarded as a long-term research objective. Ultimately, identifying the most efficient distributed architecture will require a comprehensive cost-per-bit analysis that jointly considers system performance, implementation impairments, deployment constraints, and formation-flying requirements. The main contribution of this work is therefore to demonstrate the substantial performance potential of large distributed apertures, thereby motivating future research efforts focused on overcoming the associated synchronization, positioning, scalability, and cost-efficiency challenges.

\balance

\bibliographystyle{ieeetr}
\bibliography{biblio}

\end{document}